# Single-ended Coherent Receiver

Son Thai Le, *Member, IEEE,* Vahid Aref, *Member, IEEE* and Junho Cho, *Senior Member, IEEE*

*Abstract* — **Commercial coherent receivers utilize balanced photodetectors (PDs) with high single-port rejection ratio (SPRR) to mitigate the signal-signal beat interference (SSBI) due to the square-law detection process. As the symbol rates of coherent transponders are increased to 100 Gbaud and beyond, maintaining a high SPRR in a cost-effective manner becomes more and more challenging. One potential approach for solving this problem is to leverage the concept of single-ended coherent receiver (SER) where single-ended PDs are used instead of the balanced PDs. In this case, the resulting SSBI should be mitigated in the digital domain. In this paper, we show that SSBI can be effectively mitigated using various low-complexity techniques, such as the direct filed reconstruction (DFR), clipped iterative SSBI cancellation (CIC) and gradient decent (GD). In addition, we present a self-calibration technique for SERs which can be extended for characterizing the optical-to-electrical (O/E) response of a conventional balanced coherent receiver (BR). Using the developed techniques, we then experimentally demonstrate a 90 Gbaud probabilistically constellation shaped 64-QAM (PCS-64QAM) transmission using a SER, achieving a net data rate of 882 Gb/s over 100 km of standard single mode fiber (SSMF). The sensitivity penalty compared to the BR is below 0.5 dB. We expect that when the symbol rate is increased further, a SER can potentially outperform a BR, especially when applied to cost-sensitive commercial pluggable coherent transceivers.**

## I. INTRODUCTION

T he global internet protocol (IP) and interconnect traffic have been growing exponentially with compound annual growth rates (CAGRs) of 30% and 52% respectively over the last two decades [1]. In addition, due to the ever-increasing demands for mobile broadband connectivity (e. g., 5G) and cloud services for processing and storage, it is expected that the traffic will keep growing exponentially over the next decade. This trend has put a lot of pressure on network operators, global content network (GCN) and web-scale companies in evolving their networks to support the traffic demands.

Pluggable optical transceivers have been playing a crucial role in optical networks, spanning from access, regional to mobile backhaul, mobile fronthaul and datacenter networks [2]. Compared to the traditional embedded transport solutions, pluggable transceivers offer several important advantages such as low cost, interoperability, fast deployment, and fast troubleshooting. Over the last decade, pluggable transceivers have evolved from 10 Gb/s non-return-to-zero (NRZ) modulation [3] to 100 Gb/s with 4-array pulse amplitude modulation (PAM4) [4] to 400 Gb/s polarization multiplexed quadrature amplitude modulation (PM-QAM) and coherent detection [5, 6]. Beside the increase in module's data rate, the symbol rate has also been increased continuously. Increasing the symbol rate is an effective approach to reducing the cost per transmitted bit. State-of-the-art 400 Gb/s ZR coherent transceivers operate at 59.84375 Gbaud using PM-16QAM format. For achieving the next target data rate of 800 Gb/s, 128 Gbaud PM-16QAM is being considered [7]. For future 1.6 Tb/s modules, it is possible that a symbol rate beyond 200 Gbaud will be implemented.

Commercial coherent transceivers utilize balanced photodetectors (BPDs) with high single-port rejection ratio (SPRR) to mitigate the signal-signal beat interference (SSBI) due to the square-law detection process [8, 9]. The SPRR depends on the balance (in both amplitude and skew) of the optical hybrid, RF circuity and also the similarity of the two single-ended photodetectors (PDs) of a BPD in responsivities, polarization dependencies and frequency responses [10]. As shown in [10], due to the skew, SPRR is frequency dependent as it decreases with increasing RF frequency. In a similar manner, the matching of two single-ended PDs of a BPD, which is usually characterized by the common mode rejection ratio (CMRR)) also degrades with increasing RF frequency. As a result, for a practical coherent receiver, SPRR degrades at high frequencies. Because of this, when pushing the symbol rate beyond 100 Gbaud, maintaining a high SPRR in a cost-effective manner becomes more and more challenging.

One potential approach to overcoming the abovementioned issue is to leverage the concept of single-ended coherent receiver (SER) [11-14]. In a SER, 4 single-ended PDs are used instead of 4 BPDs. One clear benefit of the SER over the conventional balanced coherent receiver (BR) is the simplicity and cost reduction because single-ended PD is cheaper than a BPD. Furthermore, as the SER does not rely on the balance of optical hybrids and the identity of PDs, the concept of SPRR becomes irrelevant. This potentially allows for relaxed fabrication requirement, higher yield, and faster testing which, in turn, will further reduce the cost. On the other hand, as the SSBI is not rejected by a SER, it should be dealt with in the digital domain. The efficiency and complexity of such SSBI removal would define when SERs can provide practical benefits over the conventional BRs.

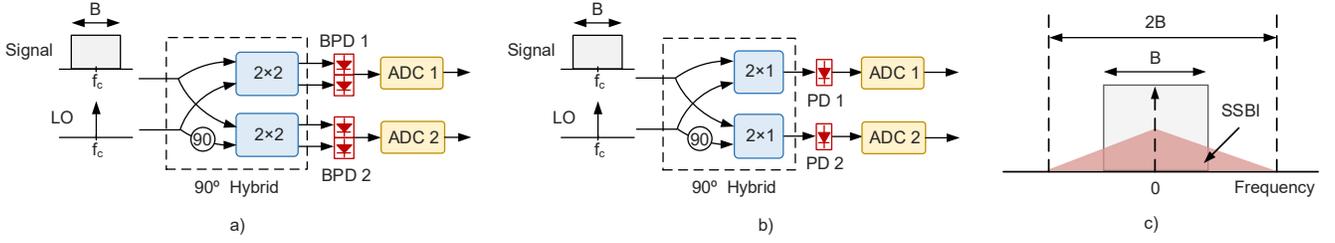

Fig. 1a) – Balanced coherent receiver in single polarization; b) – Single ended coherent receiver in single polarization; c) – Illustration of SSBI in a SER

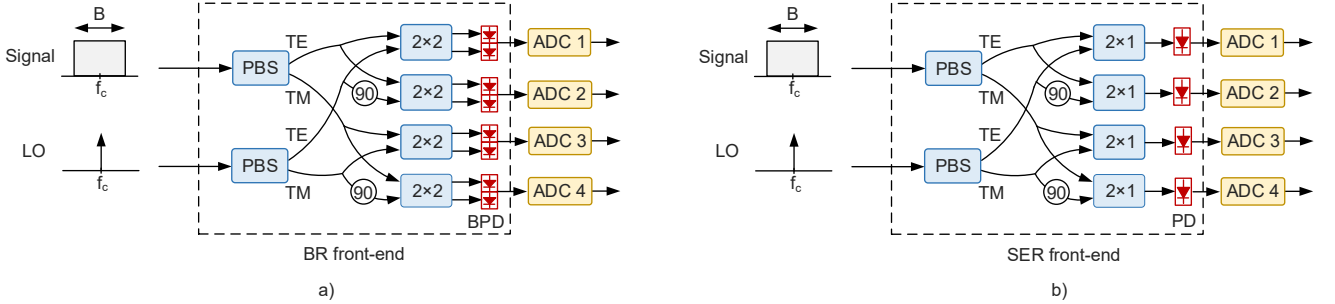

Fig. 2a) – Balanced coherent receiver in dual polarization; b) – Single ended coherent receiver in dual polarization.

In this paper, we propose and discuss three signal processing techniques for SSBI removal in SERs, namely the direct filed reconstruction (DFR), clipped iterative SSBI mitigation (CIC) and the gradient decent (GD). In addition, we propose an effective scheme for the SER front-end characterization. The developed techniques are successfully applied in a 90 Gbaud probabilistically constellation shaped 64-QAM (PCS-64QAM) transmission over 100 km, providing a net data rate of 882 Gb/s. The paper is organized as follow: Section II discusses the concept of SER and the necessity for SSBI removal, section III presents three efficient techniques for SSBI removal in SERs, section IV proposes a characterization technique for SERs and explains how this technique can be used in conventional BRs for enabling self-calibration, section V presents the experimental setup and transmission results for 90 Gbaud PCS-64QAM, and finally section VI concludes the paper.

## II. Single-ended Coherent Receiver

The block diagram of a SER in single polarization is shown in Fig. 1b in comparison with its BR counterpart (Fig. 1a). One should note that a dual-polarization SER can be simply considered as a combination of two single-polarization SERs as depicted in Fig. 2. For this reason, and for simplicity, we only consider the case of single polarization in all theoretical analysis in this paper.

As shown in Fig. 1, a single-polarization SER consists of a simplified 90º optical hybrid with two 2×1 couplers instead of two 2×2 couplers in the conventional BR (Fig. 1a). It also uses 2 single-ended PDs instead of 2 BPDs.

Let us assume that the received complex optical signal is $E(t) = \sqrt{2}(I(t) + jQ(t))$ and the local oscillator (LO) is $E_{LO}(t) = \sqrt{2}A$, where $A$ is a positive number. The usage of $\sqrt{2}$ factor is to account for the 3 dB loss of the ideal 3-dB splitter for the signal and LO paths. One can note that both the carrier frequency offset and phase noise can be included in the complex representation of the received optical signal. The signals at the inputs of two PDs are then expressed as:

$$\begin{cases} E_1(t) = I(t) + jQ(t) + A \\ E_2(t) = I(t) + jQ(t) + Ae^{j\pi/2} \end{cases} \quad (1)$$

Let us further assume that the PDs are ideal (ideal square-law detectors), then the detected analog signals at the outputs of the two single-ended PDs can be written as:

$$\begin{cases} R_1(t) = A^2 + I(t)^2 + Q(t)^2 + 2AI(t) \\ R_2(t) = A^2 + I(t)^2 + Q(t)^2 + 2AQ(t) \end{cases} \quad (2)$$

One can note that the detected signals include both the useful detection terms, $2AI(t)$ and $2AQ(t)$, and the SSBI ($I(t)^2 + Q(t)^2$) that is proportional to the detected optical signal intensity. In this case, a SER can be characterized by the signal to interference ratio (SIR) which is defined as:

$$\text{SIR} = \mathbb{E}\left\{\frac{4A^2(I(t)^2 + Q(t)^2)}{2(I(t)^2 + Q(t)^2)^2}\right\} = \mathbb{E}\left\{\frac{4A^2}{P}\right\}, \quad (3)$$

where $\mathbb{E}\{\cdot\}$ denotes expectation, and $P = 2\mathbb{E}\{I(t)^2 + Q(t)^2\}$ is the received signal power. One can verify the relation between SIR and the LO-to-signal-power ratio (LOSPR) is given as:

$$\text{SIR} = \frac{2A^2}{\mathbb{E}\{I(t)^2 + Q(t)^2\}} = 2\text{LOSPR} \quad (4)$$

As illustrated in Fig. 1c, if we assume that the signal is confined within the frequency band $[-B/2, B/2]$ Hz, where $B$ is the optical signal bandwidth, then the SSBI is confined within the frequency band $[-B, B]$ Hz. If we further assume that the signal spectrum is flat within $[-B/2, B/2]$ Hz, then the power spectrum of $I(t)^2 + Q(t)^2$ has a triangular shape within $[-B, B]$ Hz. Then, it can be easily shown that only 75% of the

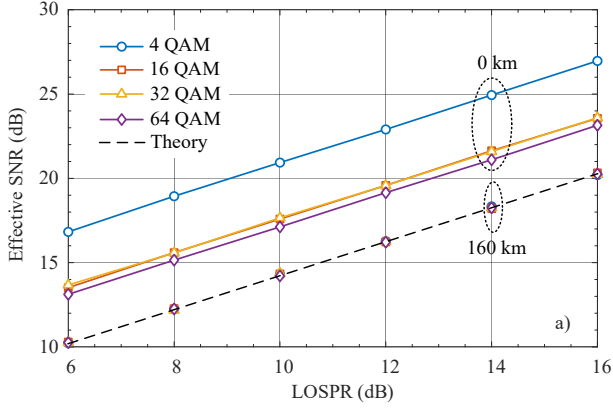 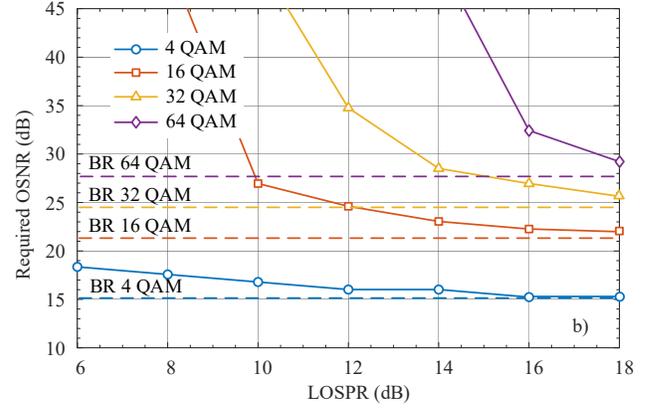

Fig. 3a) – SNR of a SER in B2B and over 160 km of SSMF in C-band for 100 Gbaud signals with different modulation formats (4 QAM, 16 QAM, 32 QAM and 64 QAM); b) – The required OSNR for 100 Gbaud signals with different modulation formats (4 QAM, 16 QAM, 32 QAM and 64 QAM). The pulse-shaping is RRC pulse with 1% roll-off.

SSBI power falls within the signal bandwidth. As a result, the in-band SIR is given as:

$$\text{In-band SIR} = \frac{8}{3}\text{LOSPR}. \quad (5)$$

For input signal having a Gaussian-like distribution, e. g. high order QAM with pulse-shaping or signals impaired by fiber chromatic dispersion (CD), the in-band SIR can be used to predict the system performance. More precisely, without additive noise from the channel, the signal-to-noise ratio (SNR) should be equal to the in-band SIR:

$$\text{SNR[dB]} \sim \text{In-band SIR[dB]} = \text{LOSPR[dB]} + 4.259. \quad (6)$$

This equation indicates an important feature of SERs – the achievable SNR is directly proportional to the LOSPR. This means that increasing the LO power by 1 dB increases the system SNR by 1 dB if the channel is noise-less.

Using simulation, we investigate the performance of a 100 Gbaud transmission system in B2B and over 160 km with a SER and various modulation formats in Fig. 3a. The effective SNR is calculated using received QAM symbols. In this simulation, we use 1% root-raise cosine (RRC) pulse shaping. One should note that the pulse shaping might have a strong impact on the B2Bperformance, as it changes the signal's peak to average power ratio (PAPR).

In Fig. 3a, we can see that in the B2B system performance strongly depends on the modulation format where 4 QAM shows ~ 4 dB advantage over 16 QAM and 32 QAM formats. On the other hand, over 160 km of transmission in C-band, all the considered modulation formats are recovered with the same effective SNR, which coincides with Eq. 6. This is because the received signal distribution approaches a Gaussian distribution due to the fiber CD regardless of the modulation format. The result presented in Fig. 3 confirms the validity of Eq. 6, showing that the transmission performance of a system with SER depends strongly on the LOSPR. At a LOSPR of 12 dB, the achievable SNR is only ~ 16 dB, which may be insufficient for high-order modulation formats.

Figure 3b shows the required optical SNR (OSNR) at a bit-error-rate (BER) of 2e-2 for the 100 Gbaud transmission system over 160 km with various modulation formats. In comparison with the ideal BR, an ONSR penalty of 1 dB can be observed at a LOSPR of ~ 12 dB for 4 QAM. For a similar OSNR penalty, a greater LOSPR of 16 dB and above 18 dB would be required for 16 QAM and 32 QAM formats. In practice, the LO power is limited and operating the receiver with a low LO power is desirable as this enhances the power efficiency of the transceiver. As a result, the conventional SER does not seem to be practical enough for commercial deployment.

## III. FIELD RECONSTRUCTION FOR SER

In the previous section, we treat the interference (SSBI) in SER simply as Gaussian-distributed noise. This approach is clearly pessimistic as the SSBI is signal dependent as can be observed in Eq. 2. In fact, Eq. 2 describes a system of 2 quadratic equations with 2 unknowns, $I(t)$ and $Q(t)$, for which a unique solution can be found analytically under certain conditions. In addition, various numerical approaches can also be used to approximately solve this equation. In this section, we propose and compare three of such approaches.

### A. Direct Field Reconstruction

One can note from Eq. 2 that the interference (SSBI) is signal dependent, which suggests that it can be removed through digital signal processing (DSP). Let us denote the outputs of PDs sampled by analog-do-digital converters (ADCs) as $R_{1,2}(k) = R_{1,2}(t = kT)$ where $T$ is the ADC's sampling period. Since $R_{1,2}(k)$ is assumed to be stationary, for notational convenience, we drop the index $k$ in all analyses below. To account for the power imbalance of the received PDs, we modify Eq. 2 as:

$$\begin{cases} R_1 = A_1^2 + I^2 + Q^2 + 2A_1 I \\ R_2 = A_2^2 + I^2 + Q^2 + 2A_2 Q \end{cases}, \quad (7)$$

where $A_1 \neq A_2$ due to the imbalances of optical hybrid and the two single-ended PDs. One should note that $R_1$ and $R_2$ have been rescaled such that the sum of the quadratic term ($I^2 + Q^2$) become equal between $R_1$ and $R_2$.

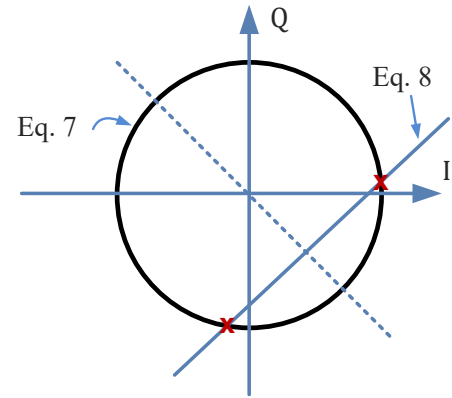

Fig. 4. Visualization of two possible solutions (crosses) of Eq. 8 and Eq. 9. The acceptable solution of Eq. 11 and Eq. 12 is the one above the dashed line

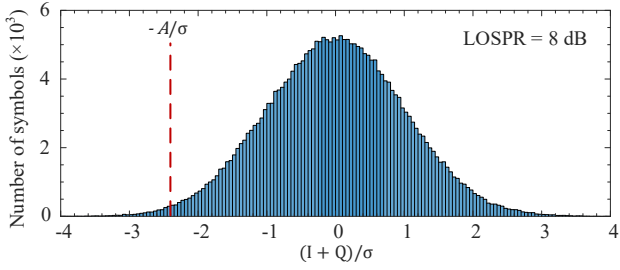

Fig. 5. The histogram of the received $I + Q$. The dash line is $-A/\sigma$ showing that if $I + Q + A < 0$, these symbols are detected in error. The DSER is $Q(2.51) \approx 6e - 3$.

From Eq. 7, one can verify that:

$$A_1 \left( I + \frac{A_1}{2} \right) - A_2 \left( Q + \frac{A_2}{2} \right) = \frac{1}{2}(R_1 - R_2) \quad (8)$$

and

$$\left( I + \frac{A_1}{2} \right)^2 + \left( Q + \frac{A_2}{2} \right)^2 = \frac{1}{2}(R_1 + R_2) - \frac{A_1^2 + A_2^2}{4} \quad (9)$$

Note that Eq. 8 characterizes a line in 2D space and Eq. 9 characterizes a circle. A line and a circle intersect on either no points, a single point or 2 points as illustrated in Fig. 4. This means that the solution may not be unique.

For notational convenience, let us define:

$$A_x = \sqrt{A_1^2 + A_2^2},$$

$$\Delta = 4R_1 R_2 - (R_1 + R_2 - A_x^2). \quad (10)$$

Then one can verify that

$$I = -\frac{A_1}{2} + \frac{A_1}{2A_x^2}(R_1 - R_2) \pm \frac{A_2}{2A_x^2}\sqrt{\Delta}, \quad (11)$$

$$Q = -\frac{A_2}{2} - \frac{A_2}{2A_x^2}(R_1 - R_2) \pm \frac{A_1}{2A_x^2}\sqrt{\Delta}. \quad (12)$$

Equations (11) and (12) form the basic principle of the direct field reconstruction (DFR) technique for SERs.

For the special case when $A_1 = A_2 = A$ (no imbalance), we have:

$$\Delta = 4R_1 R_2 - (R_1 + R_2 - 2A^2)^2, \quad (13)$$

and

$$I = -\frac{A}{2} + \frac{1}{4A}(R_1 - R_2) \pm \frac{1}{4A}\sqrt{\Delta}, \quad (14)$$

$$Q = -\frac{A}{2} - \frac{1}{4A}(R_1 - R_2) \pm \frac{1}{4A}\sqrt{\Delta}. \quad (15)$$

One can verify that $\Delta$ in Eq. (13) can also be written as:

$$\Delta = 4A^2(I + Q + A)^2, \quad (16)$$

which means that $\Delta \geq 0$ and at least one solution for $(I, Q)$ exists. Then, summing Eq. (15) and (16) results in:

$$I + Q + A = \pm \frac{1}{2A}\sqrt{\Delta}. \quad (17)$$

From this equation, one can note that the uniqueness of solution depends on the sign of $I + Q + A$. If the LO power is sufficiently large such that $I + Q + A \geq 0$, then a unique solution (outputs of DSP) is given by:

$$\hat{I} = -\frac{A}{2} + \frac{1}{4A}(R_1 - R_2) + \frac{1}{4A}\sqrt{|\Delta|}, \quad (18)$$

$$\hat{Q} = -\frac{A}{2} - \frac{1}{4A}(R_1 - R_2) + \frac{1}{4A}\sqrt{|\Delta|}. \quad (19)$$

Here, note that a modulus operator ($|\cdot|$) is included, since in practice $\Delta$ can have small negative values due to the Rx noise and its O/E front-end response, especially at a low LOSPR.

In certain practical implementations, the condition $I + Q + A \geq 0$ might not be satisfied all the time; for example, when the LOSPR is low, the DFR technique based on Eq. 18 and Eq. 19 would be in error. The detector symbol error rate (DSER) is then defined as:

$$\text{DSER} = Pr\{I + Q + A < 0\} \quad (20)$$

The samples $I$ and $Q$ can be modeled as independent and identically distributed random variables whose statistical distribution typically depends on the modulation format, pulse-shaping, and the CD. For high-baudrate transmissions over datacenter interconnect (DCI) or longer distances, the samples can be modelled as Gaussian random variables as:

$$I \sim \mathcal{N}\left( 0, \frac{P}{4} \right), \qquad Q \sim \mathcal{N}\left( 0, \frac{P}{4} \right). \quad (21)$$

In this case, one can easily verify that.

$$\text{DSER} = \mathcal{Q}\left( \frac{A}{\sigma} \right) = \mathcal{Q}(\sqrt{LOSPR}), \quad (22)$$

where $\mathcal{Q}$ is the Q-function (e. g., $\mathcal{Q}(x) = \int_x^\infty \exp(-x^2/2) dx / \sqrt{2\pi}$) and $\sigma^2 = P/2$ is the variance of $I + Q$. The samples which are detected in error are illustrated in Fig. 5 for LOSPR = 8 dB, showing a DSER $\sim \mathcal{Q}(2.51) \approx 6e - 3$.

The mean square error (MSE) of the detector is defined as:

$$\text{MSE} = \frac{\mathbb{E}\left\{ \left( \hat{I} - I \right)^2 + \left( \hat{Q} - Q \right)^2 \right\}}{\sigma^2}. \quad (23)$$

From Eq. 17-19, one can verify

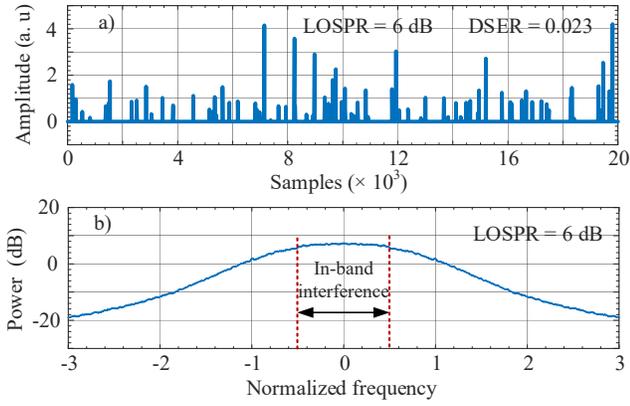

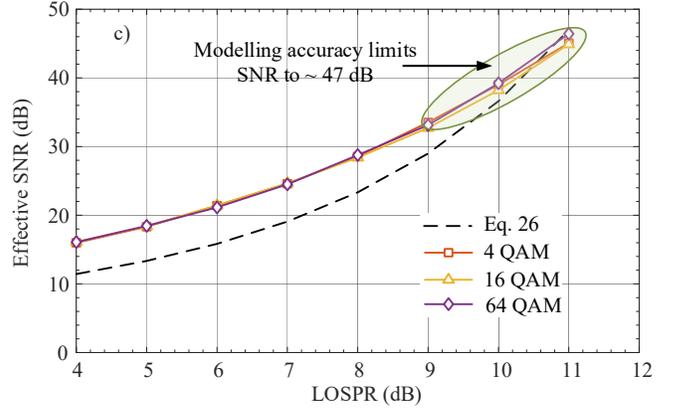

Fig. 6a – Reconstruction error $(\hat{I} - I)$ (a) and its power spectrum density (b) for DFR technique with 100 Gbaud 64 QAM signal over 160 km at LOSPR of 6 dB. The two dash lines in Fig. 6b show the in-band interference. c) –SNR for SER with DFR with 100 Gbaud signals over 160 km

$$\left(\hat{I} - I\right)^2 + \left(\hat{Q} - Q\right)^2 = 2(I + Q + A)^2, \quad (24)$$

which leads to

$$\begin{aligned}
\text{MSE} &= \frac{2}{\sigma^3 \sqrt{2\pi}} \int_{-\infty}^{-A} (z + A)^2 \exp\left(-\frac{z^2}{2\sigma^2}\right) dz \\
&= \frac{2}{\sqrt{2\pi}} \int_{-\infty}^{-A/\sigma} (z + A/\sigma)^2 \exp(-z^2/2) \, dz \\
&= 2\left(1 + \frac{A^2}{\sigma^2}\right) Q\left(\frac{A}{\sigma}\right) - \frac{2A}{\sigma\sqrt{2\pi}} \exp\left(-\frac{A^2}{2\sigma^2}\right). \quad (25)
\end{aligned}$$

The SNR can be estimated from the MSE as:

$$\text{SNR [dB]} = -10\log_{10}(\text{MSE}), \quad (26)$$

Figure 6a shows the reconstruction error $\left(\hat{I} - I\right)$ for 100 Gbaud 64 QAM signal over 160 km when the LOSPR is set to 6 dB. One can note the discrete nature of $\hat{I} - I$ as $\hat{I} - I \neq 0$ only at samples where $I + Q + A < 0$. The power spectrum density of $\hat{I} - I$ is shown in Fig. 6b where the frequency is normalized to the baudrate, e. g., 100 Gbaud. Due to the discrete nature of $\hat{I} - I$, its spectrum is significantly wider than the signal's bandwidth. As a result, a low pass filter can significantly suppress the impact of reconstruction error. In addition, the reconstruction errors are at high amplitude samples, which usually are not critical for QAM symbol detections. As a result, we expect that the theoretical estimation based on Eq. (26) underestimates the true performance of DFR technique.

In Fig. 6c, we plot the SER's SNR as a function of LOSPR for 100 Gbaud 64 QAM signal over 160 km. One can note that, in the presence of CD (160 km of distance), the DFR technique shows the same performance for 4 QAM, 16 QAM and 64 QAM formats. As explained above, Eq. 26 underestimates the system performance by ∼ 5 dB when LOSPR is below 8 dB. At a relatively low LOSPR value of 10 dB, DFR shows an excellent SNR of ∼ 39 dB. We note that at this high SNR, the simulation result is strongly affected by the modelling accuracy, e, g., by the use of pulse-shaping filter truncation and double precision floating point, which limits the SNR to ∼ 47 dB. We suspect that the crossing point between simulation and Eq. 26 at LOSPR of ∼ 11 dB is due to the modelling accuracy. Overall, Fig. 6c shows the exceptional performance of the DFR, where even a LOSPR of 6 dB already leads to a SNR above 20 dB and a LOSPR of 11 dB leads to a SNR beyond 45 dB, at which the SER can be well considered as a linear receiver.

### B. Iterative SSBI cancellation

Let us consider a special case of Eq. (6) when $A_1 = A_2 = A$ and develop an iterative SSBI cancellation (IC) scheme for it. The general case when $A_1 \neq A_2$ can be treated in a similar manner. For convenience, we make the change of variables as:

$$U_1 = \frac{R_1 - A^2}{4A^2}; U_2 = \frac{R_2 - A^2}{4A^2}; \ \bar{I} = \frac{I}{2A} \text{ and } \overline{Q} = \frac{Q}{2A},$$

then one can verify

$$\begin{cases} U_1 = \bar{I}^2 + \overline{Q}^2 + \bar{I} \\ U_2 = \bar{I}^2 + \overline{Q}^2 + \overline{Q} \end{cases}. \quad (27)$$

By assuming that the LOSPR is sufficient and that the SSBI can be considered relatively small compared to $2\bar{I}$ and $2\overline{Q}$, we can choose the initial guess for the solution of Eq. 27 as:

$$\bar{I}_{(0)} = U_1 - \frac{P}{4A^2}; \quad \overline{Q}_{(0)} = U_2 - \frac{P}{4A^2}, \quad (28)$$

where $\mathbb{E}\{\bar{I}_{(0)}\} = \mathbb{E}\{\overline{Q}_{(0)}\} = 0$ since $\bar{I}$ and $\overline{Q}$ have zero means based on the assumption (21)

The estimation error at this stage is calculated as:

$$\Delta \bar{I}_{(0)} = \bar{I}_{(0)} - \bar{I} = \bar{I}^2 + \overline{Q}^2 - \frac{P}{4A^2}. \quad (29)$$

One then can determine the SSBI and reduce the estimation errors in following iterations as [12]:

$$\begin{cases} \bar{I}_{(n+1)} = U_1 - \left(\bar{I}_{(n)}^2 + \overline{Q}_{(n)}^2\right) \\ \overline{Q}_{(n+1)} = U_2 - \left(\bar{I}_{(n)}^2 + \overline{Q}_{(n)}^2\right) \end{cases}. \quad (30)$$

One can then easily see that the estimation errors have the recurrence relation:

**Table 1.** Convergence behavior of the iterative SSBI cancellation scheme (31)

1) If (34) is satisfied without the "=" sign, the algorithm diverges as $\Delta \overline{I}_{(n)} \rightarrow -\infty$.

2) If (34) is satisfied with the "=" sign, there are 2 scenarios:

   a. If $\overline{I} + \overline{Q} \leq -A$ then $\Delta \overline{I}_{(n)}$ converges to 0.

   b. If $\overline{I} + \overline{Q} > -A$ then $\Delta \overline{I}_{(n)}$ converges to $-\left(\overline{I} + \overline{Q} + 0.5\right)$.

3) If (34) is not satisfied, there are 4 scenarios:

   a. If $|\overline{I} + \overline{Q}| \leq A$ then $\Delta \overline{I}_{(n)}$ converges to 0.

   b. If $-3A \leq \overline{I} + \overline{Q} < -A$, then $\Delta \overline{I}_{(n)}$ converges to $-\left(\overline{I} + \overline{Q} + 0.5\right)$.

   c. If $-4A \leq \overline{I} + \overline{Q} < -3A$ or $A < \overline{I} + \overline{Q} \leq 2A$ then $\Delta \overline{I}_{(n)}$ bounces between multiple values and can even go to a chaotic regime.

   d. If $\overline{I} + \overline{Q} < -4A$ or $\overline{I} + \overline{Q} > 2A$ the algorithm does not converge, and $\Delta \overline{I}_{(n)}$ can either be bounded in some region or go to $-\infty$.

---

$$\Delta \overline{I}_{(n+1)} = -\Delta \overline{I}_{(n)} \left(\overline{I} + \overline{Q} + \overline{I}_{(n)} + \overline{Q}_{(n)}\right)$$
$$= -2\Delta \overline{I}_{(n)}\left(\Delta \overline{I}_{(n)} + \overline{I} + \overline{Q}\right), \quad (31)$$

and we always have:

$$\Delta \overline{Q}_{(n)} = \Delta \overline{I}_{(n)}, \quad (32)$$

The success of the IC algorithm (30) in cancelling the SSBI implies

$$\left|\Delta \overline{I}_{(n)}\right| \rightarrow 0 \text{ when } n \rightarrow \infty. \quad (33)$$

In this paper, we analyze the convergence behavior of the recurrence sequence (31) for the first time in literature. Fig. 7 shows the bifurcation diagram [15-17] of $\Delta \overline{I}_{(1000)}$, which shows the possible values that $\Delta \overline{I}_{(n)}$ can have after a large number of iterations $n = 1000$, obtained with Monte-Carlo simulation using random $(\overline{I}, \overline{Q})$ samples that satisfies the following initial condition:

$$\Delta \overline{I}_{(0)} + \overline{I} + \overline{Q} \leq \frac{1}{2} + \left|\overline{I} + \overline{Q} + \frac{1}{2}\right|. \quad (34)$$

One can see from Fig. 7 that depending on the initial values of $(\overline{I}, \overline{Q})$, the estimation error (31) can converge to 0 (implying that the IC (30) completely removes the SSBI) or a non-zero fixed value (implying that the IC converges, but to a wrong solution), or it can have multiple values (the estimation error in this case bounces within a fixed interval over iterations). The convergence behavior of (31) under various conditions is summarized in Tab. 1, and details of how it is obtained are given in Appendix I.

From Fig. 7 and Tab. 1, one can note that if $|\overline{I} + \overline{Q}| \leq A$ and (34) is satisfied, then $\Delta \overline{I}_{(n)}$ converges to 0, which implies that the IC algorithm (30) can successfully remove the SSBI. Otherwise, the IC algorithm (30) will result in estimation error. If we assume Gaussian distributions for both $I$ and $Q$ as shown in (21), and using the same argument therein, the estimation symbol error rate (ESER) of the IC algorithm (3) can be estimated as:

$$\text{ESER} \approx 2Q\left(\frac{A}{\sigma}\right) = 2Q(\sqrt{LOSPR}), \quad (35)$$

which is approximately twice as much as the detector symbol error rates of the DFR technique (Eq. 22). For LOSPR = 8 dB,

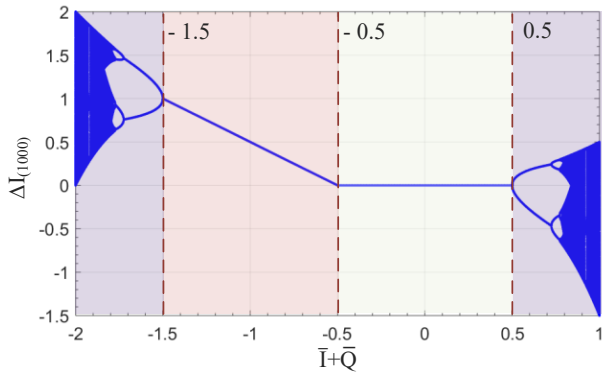

Fig. 7. Bifurcation diagram of $\Delta \overline{I}_{(1000)}$ versus $\overline{I} + \overline{Q}$, obtained from $10^5$ random values of $\Delta \overline{I}_{(0)}$ for each $\overline{I} + \overline{Q}$ which satisfies the condition (34)

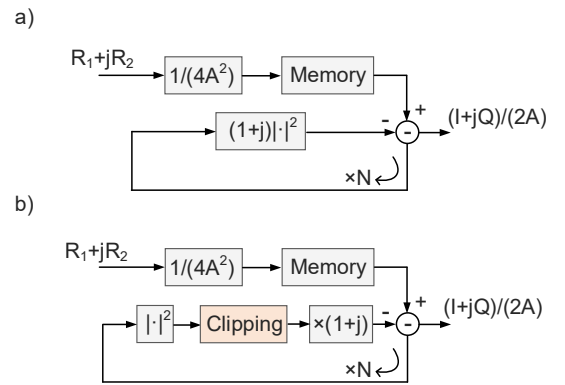

Fig. 8. Block diagrams of iterative SSBI cancellation without (a) and with clipping (b)

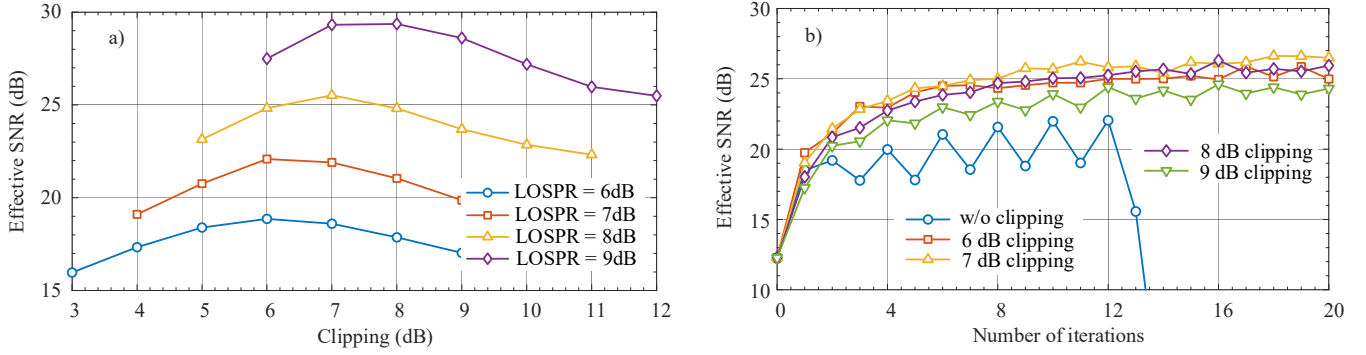

Fig. 9a) – Optimizing the clipping level in dB for various values of LOSPR for CIC with 12 iterations; b) – Performance versus number of iterations for LOSPR = 8 dB. The system under test is 100 Gbaud 64 QAM transmission over 160 km.

the ESER of the IC algorithm (30) is estimated to be ∼ $2Q(2.51) \approx 1.2e - 2$.

When $|I + Q| > A$, the IC algorithm (30) either diverges or results in a finite non-zero estimation error. This indicates that the IC algorithm (30) might not be effective for high-peak power samples. For such samples, the IC algorithm can actually enhance the SSBI instead of reducing it. When the number of iterations increases, the enhanced SSBI can even go to infinity, causing numerical instability. Fortunately, these samples can be detected by monitoring the peak-to-average power (PAPR) of the output signal after each iteration. For practical implementation, a clipper can be included in the iterative loop as shown in Fig. 8 [18]. We call this scheme *clipped iterative SSBI cancellation* (CIC). This idea has been shown to be also very effective for SSBI mitigation in single-sideband (SSB) transmission [19-23].

For the CIC technique, choosing an appropriate clipping level is an important task. As the clipping is applied on estimated SSBI which is an intensity signal, we can define the clipping level expressed in dB relative to the average signal power. Fig. 10a depicts the effective SNR versus the clipping level for 100 Gbaud 64 QAM transmission over 160 km using the CIC technique with 12 iterations. One can note that for each value of LOSPR, there is an optimum clipping level, which is around 1 dB less than the LOSPR. A comparison of the IC with and without clipping is shown in Fig. 10b for LOSPR = 8 dB. One can note that when clipping is not applied, numerical instability occurs after 12 iterations, causing catastrophic performance degradation. This is due to the exponential growth of the reconstruction errors on samples with large amplitudes as discussed before. Overall, an SNR of ∼ 21 dB can be achieved when clipping is not applied. On the other hand, when clipping is applied, continuous performance improvement can be achieved when increasing the number of iterations. A SNR of ∼ 26 dB can be achieved with optimum clipping level (7dB), which shows a significant benefit of the proposed clipping technique.

### C. Gradient decent

For solving Eq. 27, the well-known gradient decent (GD) [24] method can be used. For convenience, we define:

$$\begin{cases} X(\overline{I}, \overline{Q}) = \overline{I}^2 + \overline{Q}^2 + \overline{I} - U_1 \\ Y(\overline{I}, \overline{Q}) = \overline{I}^2 + \overline{Q}^2 + \overline{Q} - U_2 \end{cases}, \quad (36)$$

Then Eq. 27 becomes

$$\begin{cases} X(\overline{I}, \overline{Q}) = 0 \\ Y(\overline{I}, \overline{Q}) = 0 \end{cases}. \quad (37)$$

To solve Eq. 36, we define the objective function as

$$G(\overline{I}, \overline{Q}) = X(\overline{I}, \overline{Q})^2 + Y(\overline{I}, \overline{Q})^2, \quad (38)$$

which we will attempt to minimize.

The initial guess can be chosen as in Eq. 28. The GD update rules for minimizing $G(\overline{I}, \overline{Q})$ can be expressed as:

$$\begin{aligned} \overline{I}_{(n+1)} = \overline{I}_{(n)} - \mu \big[ & X\left(\overline{I}_{(n)}, \overline{Q}_{(n)}\right)\left(2\overline{I}_{(n)} + 1\right) \\ & + 2Y\left(\overline{I}_{(n)}, \overline{Q}_{(n)}\right)\overline{I}_{(n)} \big], \end{aligned}$$

$$(39)$$

$$\begin{aligned} \overline{Q}_{(n+1)} = \overline{Q}_{(n)} - \mu \big[ & X\left(\overline{I}_{(n)}, \overline{Q}_{(n)}\right)\overline{Q}_{(n)} \\ & + 2Y\left(\overline{I}_{(n)}, \overline{Q}_{(n)}\right)\left(2\overline{Q}_{(n)} + 1\right) \big], \end{aligned}$$

$$(40)$$

where $\mu$ is the step size.

To study the convergence behavior of the GD algorithm we also define the normalized error as:

$$\Delta\overline{V}_{(n)} = 2\sqrt{\left(\overline{I}_{(n)} - \overline{I}\right)^2 + \left(\overline{Q}_{(n)} - \overline{Q}\right)^2}. \quad (41)$$

The convergence behavior of the GD method is studied in Fig. 10a for a 100 Gbaud 64 QAM system over 160 km. In Fig. 10a, each line represents one sample with a different initial value $(\overline{I}, \overline{Q})$.

The convergence of the GD algorithm strongly depends on $\Delta\overline{V}_{(0)}$ as observed in Fig. 10a. When $\Delta\overline{V}_{(0)} < 1$ we can expect that the error is suppressed as the number of iterations increases. In general, the lower $\Delta\overline{V}_{(0)}$ is, the faster the conversion speed. On the other hand, when $\Delta\overline{V}_{(0)} > 1$, Fig. 10a shows that the GD algorithm might not converge correctly as the error is not suppressed when the number of iterations is increased. This observation also suggests that clipping can improve the performance of GD method. Herein, two clippers should be used for $I$ and $Q$ branches.

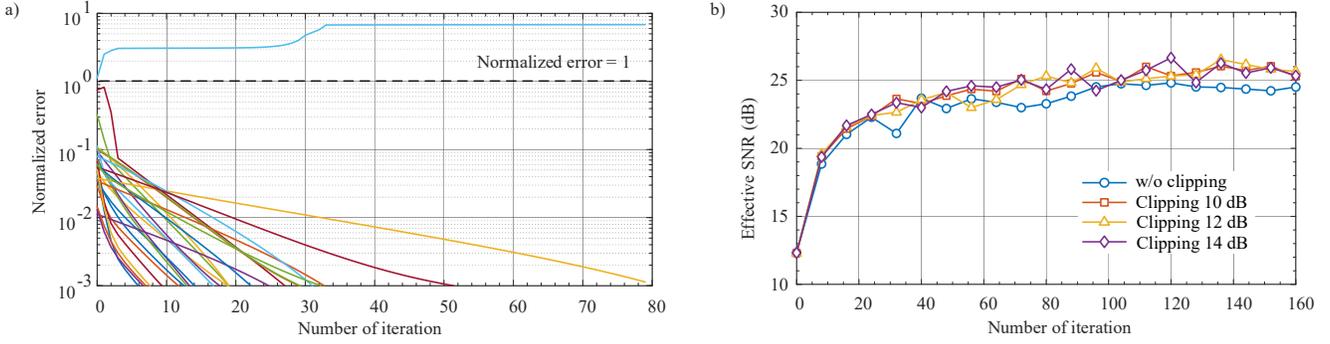

Fig. 10a) – Convergence behavior of the GD method where each line shows error evolution for one sample. Errors on samples with $\Delta \overline{v}_{(0)} > 1$ are not suppressed by the GD method; b) – Performance of the GD technique in SSBI mitigation, with and without clipping; The system under test is a 100 Gbaud 64 QAM over 160 km with LOSPR is 8 dB

Table 2. Number of real-valued multiplications required for each (I, Q) sample of DFR, CIC and GD techniques

| DFR | CIC | GD |
|-----|-----|-----|
| 10  | $2N+2$ | $6N+2$ |

The effective SNR achieved by the GD technique for SSBI mitigation in SER, with and without clipping, is shown in Fig. 10b for 100 Gbaud 64 QAM transmission over 160 km with LOSPR of 8 dB. The clipping level here is defined relative to the average power of the signal (separately for $I$ and $Q$). One can note that by applying clipping (at a clipping level of 12 dB) the SNR can be significantly improved (by more than 1 dB). The achievable SNR after 160 iterations is $\sim 26$ dB which is comparable with the CIC technique. In addition, the GD technique shows a quite slow convergence where the optimum performance can only be achieved after $\sim 100$ iterations. We note that the step size has been optimized for achieving the fast convergence. The optimum step size was $\sim 0.05$.

### D. Comparison of DFR, CIC and GD

In the previous 3 Subsections, we have introduced 3 different techniques with different levels of generality for SSBI mitigation in SER. The DFR is applicable specifically for optical filed reconstruction in SER. On the other hand, the iterative SSBI mitigation technique can be applied for SSB DD transmission and other types of DD systems as well. The GD method can be applied for solving a wide range of nonlinear equations, not just quadratic equations. The generality of the technique comes with a trade-off in performance and/or implementation complexity. The complexities of the DFR, CIC and GD techniques measured by the number of real-valued multiplications for each $\left( \overline{I}, \overline{Q} \right)$ sample are shown in Tab. 2. The DFR technique requires only 10 real-valued multiplications per sample (we assume that square root operation has a complexity of 4 real-valued multiplications). This complexity would be negligible compared to other blocks within the coherent DSP such as CD compensation, digital filtering, and phase noise compensations. This indicates that SSBI mitigation in a SER should not significantly increase the overall DSP complexity.

Unlike the DFR technique with a fixed complexity, the complexities of the CIC and GD depend on the number of iterations. When the number of iterations is small ($N < 5$) the CIC technique can be less complex than the DFR technique. For achieving the best performance, the required number of iterations can be bigger than 5, for which the CIC technique becomes more computationally expensive than the DFR technique. As discussed in the previous subsection, the GD technique converges quite slowly, which results in the GD being far the most costly technique.

In this subsection we will compare in greater detail the performance of these three techniques under some important practical constraints, namely the Rx bandwidth limitation and amplified spontaneous emission (ASE) noise. Herein, to focus on the achievable performance without implementation complexity constraints, we consider the CIC technique with 20 iterations and the GD technique with 120 iterations using the optimum clipping levels.

As depicted in Fig. 1c and Fig. 11a, the SSBI has twice the bandwidth (2B) in comparison to the signal's bandwidth (B). As a result, the SER should have at least twice the bandwidth compared to the signal's bandwidth so the signal can be detected with no loss of information. However, this requirement is not desirable and typically it cannot be met in practice. When the SER's bandwidth (denoted as $2B_{SER}$) is smaller than 2B, a portion of the SSBI is lost after the O/E conversion. This will have impact on the accuracy of SSBI mitigation schemes discussed above. To quantify the SER bandwidth, we define a bandwidth ratio (BWR) as:

$$\text{BWR} = \frac{2B_{SER}}{B}. \qquad (42)$$

Also, in simulation we model the SER's bandwidth limitation by using a brick-wall filter with normalized electrical bandwidth of BWR/2.

Figure 11b shows the comparisons of DFR, CIC and GD techniques for BWR = 2 (solid lines) and BWR = 1.2 (dashed lines). For the CIC and GD technique, we optimize the clipping ratio for each case for achieving the best performance. When BWR = 2, the optimum clipping level for CIC is $\sim \text{LOSPR} - 1$ dB, while for BWR = 1.2 the optimum clipping level is $\sim \text{LOSPR} - 2$ dB. For the GD technique, the optimum clipping level is LOSPR + 4 dB for both cases.

The case of BWR = 2 implies that the SER has enough bandwidth to capture all the SSBI. In this case, due to the exact field reconstruction principle, DFR shows the best performance for all considered values of LOSPR from 5 dB to 11 dB. GD is the worst performing technique. At low LOSPR values (below 7 dB) GD shows a comparable performance as the CIC technique. At a LOSPR of 10 dB, it provides ~ 1.5 dB and ~ 3 dB smaller effective SNR than the DFR and CIC techniques, respectively. However, for a practical BWR value of 1.2, an opposite trend is observed where GD shows the best performance, for all considered LOSPR values. This interesting result indicates that GD is less sensitive to the SER's bandwidth limitation compared to the CIC and especially the DFR technique. At LOSPR value of 9 dB, GD shows ~ 2.5 dB and 4 dB advantages over the CIC and DFR techniques.

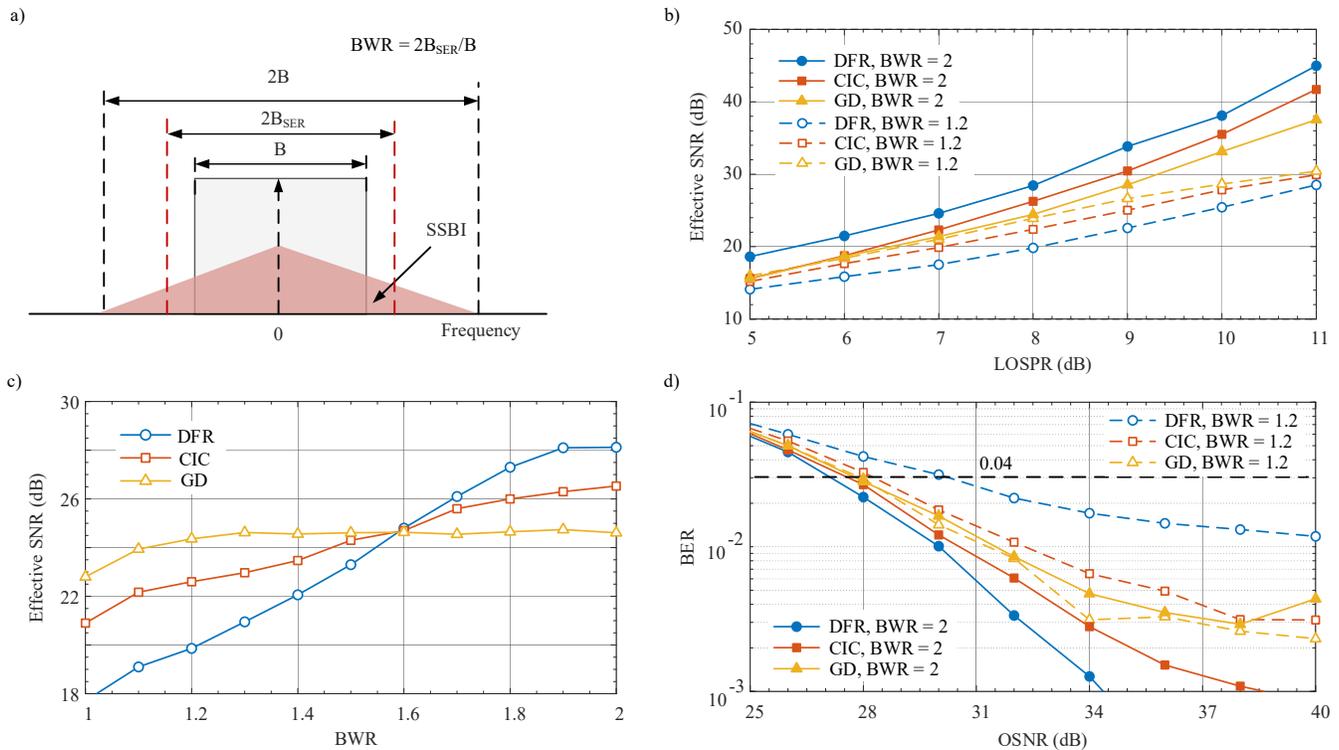

Fig. 11a) – Illustration of the Rx bandwidth limitation in a SER and the definition of BWR; b) – Comparison of DFR, CIC and GD techniques in the noise-less case for BWR = 2 and BWR = 1.2; c) – Comparison of DFR, CIC and GD techniques when the BWR is varied from 1 to 2, the LOSPR = 8 dB ; d) – ONSR performances for DFR, CIC and GD techniques with BWR of 2 and 1.2 for LOSPR = 8 dB; The system is 100 Gbaud 64 QAM over 160 km.

The impact of SER's bandwidth limitation when the LOSPR is fixed to 8 dB is shown in Fig. 11b for DFR, CIC and GD techniques. One can note that at low BWR, GD significantly outperforms the CIC and DFR technique. At a BWR of 1.4, the GD already achieve its best performance while the DFR scheme keeps performs better when the BWR is increased up to 2. At a BWR of ~ 1.6 three techniques show similar performances and after that DFR becomes the best performing technique. The key message here is that under a severe SER's bandwidth limitation, GD and CIC perform better than the DFR technique and when the SER has sufficient bandwidth the DFR would be the best technique.

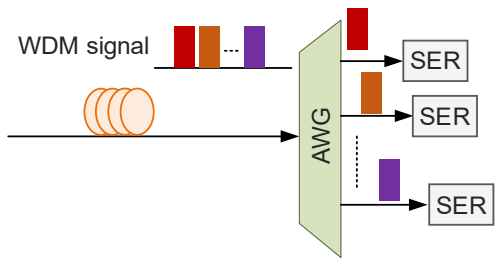

Fig. 12. SER in a WDM configuration

The BER performances of SER as a function of the OSNR are compared in Fig. 11d, between the DFR, CIC and GD techniques. As discussed, all these three techniques show excellent performance which suggests that the ASE noise is not be enhanced by the SER. In this case, the residual SSBI after SSBI mitigation can be considered as an additive noise which causes the difference in the performances of the SER with DFR, CIC and GR techniques. When BWR is set to 2, DFR shows the best OSNR performance due to its excellent SSBI suppression. On the other hand, when the BWR is set to 1.2, GD technique shows the best performance, again, due to the best suppression of the SSBI in this case. Remarkably, for the GD technique similar performances are observed for BWR = 2 and BWR = 1.2. This confirms the excellent tolerance of the GD to SER's bandwidth limitation.

Compared to the GD, the CIC technique shows only a slight performance penalty at BWR = 1.2. At the same BWR value, CIC significantly outperforms the DFR technique with ~ 2 dB OSNR advantage at the BER threshold of 0.04. This shows that CIC offers a great balance in term of performance and implementation complexity. These features could make the CIC the preferred technique for practical implementations.

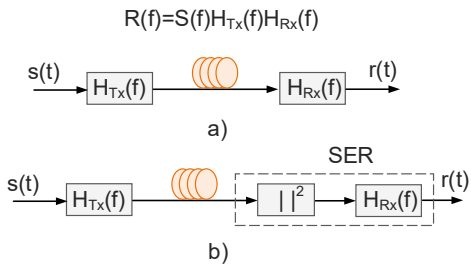

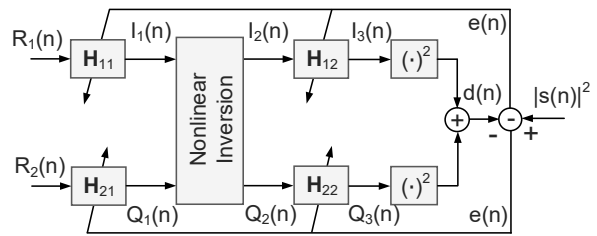

Fig. 13. Model of a linear transmission system (the Tx, Rx and channel are linear) (a) and an illustrative model for transmission system with SER (b) showing the Rx nonlinear behavior

Fig. 14. General block diagram of a digital circuit for measuring both the Tx and Rx responses of a transmission system with a SER

### E. Impact of co-propagating WDM channels

In a SER, the SSBI is mitigated in the digital domain which requires that the receiver has access to the interference signals. In the case of a WDM transmission system in the colorless reception mode [25], a SER suffers from the SSBI coming from other co-propagating channels, but it does not receive the actual signal signals from these channels. In this case, removing the SSBI coming from co-propagating channels is an impossible task. So, a SER should not be operated in the colorless mode. It requires an optical filter or demultiplexer for rejecting co-propagating channels before the channel of interest is received by a SER as shown in Fig. 12. Fortunately, most of DCI systems operate in this configuration. On the other hand, if joint processing of WDM channels is possible, the SSBI should also be removed in a similar manner as discussed above. However, this issue is out of the scope of this paper and will not be discussed further.

## IV. Rx Characterization

Results and analyses in the previous section has shown that the SSBI in a SER can be effectively mitigated using various techniques. These analyses, however, are based on an assumption that the O/E front-end response of the SER is ideal. This is not true in practice, especially for high baudrate transmissions where maintaining a flat response is very challenging and often not possible. Nonideal O/E response introduces inter-symbol interference which destroys the perfect relation of the linear detection term and the SSBI which makes the SSBI removal so effective. To explain this problem more clearly, we should include the O/E front-end responses of the SER into the Eq. 2 as:

$$\begin{cases} R_1 = J_1 \otimes (A^2 + I^2 + Q^2 + 2AI) \\ R_2 = J_2 \otimes (A^2 + I^2 + Q^2 + 2AQ) \end{cases} \tag{43}$$

where $J_1$ and $J_2$ are the impulse responses of the SER including PD, TIA and ADC and $\otimes$ denotes the convolution operator.

From Eq. 56, it is clear that the first step for retrieving $(I, Q)$ from the detected photocurrents $(R_1, R_2)$ is to determine and then reverse the SER impulse responses $J_1$ and $J_2$. In general, this can be done with a separate Rx characterization task with additional wideband characterization instrument such as vector network analyzer (VNA), wideband transmitter or tunable laser source. For achieving a good suppression of SSBI, this characterization task should be done with a high accuracy and thus it might become time consuming and/or expensive. One should note that in the conventional coherent receiver (e. g. BR), the Rx response can be corrected through DSP in the process of channel equalization. Thus, no separate Rx calibration task is required for a BR. This difference is crucial as performing a time-consuming Rx characterization task can be the showstopper for commercialization of the SER.

In order to avoid performing a separate calibration task for each SER, self-calibration techniques for SER are strongly desirable. Herein, self-calibration means that the Rx response is measured or estimated either when the system is initialized (in the B2B, loopback or the full transmission modes) or adaptively when the system is operating. This suggests that a training sequence sent from the Tx to the Rx is required.

In a fully linear transmission system (where the Tx, channel and Rx are linear) the system response can be well-obtained by using a straining sequence from the Tx (as illustrated in Fig. 13a). The conventional coherent transmission system with BR in the loopback or B2B modes can be considered as linear system. In this case, one can only measure the combined responses of the Tx and Rx by sending a training sequence. The receiver response can only be obtained if the Tx response is known or ideal. This is usually not the case. This effectively means that measuring simultaneously the Tx and Rx responses in a linear transmission system using a training sequence is not a possible task.

A SER, on another hand, is a nonlinear receiver. An illustrative model for a transmission system with a SER is depicted in Fig. 13b, showing that the Tx and Rx responses are separated by a nonlinear block. Because of this nonlinear block, the impacts of the Tx and Rx responses on the overall system's response become different. If the nonlinear block is invertible, the Tx and Rx responses can be separated. This is a major advantage of a nonlinear system over linear transmission systems which has not been widely exploited in the literature.

As we discussed in the previous section, the nonlinearity in a SER can be inverted exactly (under some conditions) using the DFR technique or approximately using the iterative SSBI cancellation scheme. Based on this feature, we can design a digital circuit which attempts to measure both the Tx and Rx responses of a SER as shown in Fig. 14. It consists of 4 FIR filters, namely $H_{11}, H_{12}, H_{21}, H_{22}$ and a nonlinear inversion block which inverts the nonlinearity of the SER. The objective of this circuit is to adaptively update the coefficients of the 4 FIR filters to minimize the following cost function:

$$C = \langle |e(n)|^2 \rangle = \langle ||s(n)|^2 - d(n)|^2 \rangle, \tag{44}$$

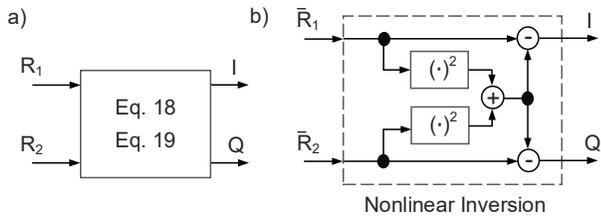

Fig. 15. Examples of nonlinear inversion blocks using a) – The DFR technique and b) – Iterative SSBI cancellation with 1 iteration

where $s(n)$ is the training sequence (e. g. QPSK or 16 QAM). One should note that the laser phase noise and carrier frequency offset have no impacts on this cost function $C$.

If we assume that the frequency offset is negligible, the cost function $C$ should be minimized when $H_{11} \otimes J_1 = H_{21} \otimes J_2 = \delta(n)$ and $H_{12} \otimes D_1 = H_{22} \otimes D_2 = \delta(n)$ where $D_1$ and $D_2$ are the corresponding Tx responses. Thus, after the algorithm's convergence, the responses of Tx and Rx can be determined.

One important design element of the proposed circuit is choosing the nonlinear inversion block as it would define both the achievable performance and implementation complexity. One obvious option (and probably the best option) is to employ Eq. 18 and Eq. 19 as illustrated in Fig. 15a. Another possible option is to employ the IC technique with only one iteration as follow:

$$\begin{cases} \hat{I}(k) = \bar{R}_1 - (\bar{R}_1^2 + \bar{R}_2^2) \\ \hat{Q}(k) = \bar{R}_2 - (\bar{R}_1^2 + \bar{R}_2^2) \end{cases}, \quad (45)$$

where:

$$\begin{cases} \bar{R}_1 = (R_1 - A^2)/(2A) \\ \bar{R}_2 = (R_2 - A^2)/(2A) \end{cases},$$

The block diagram of the nonlinear inversion block based on Eq. (58) is shown in Fig. 15b. Using a more complicated nonlinear inversion block based on iterative SSBI cancellation scheme with 2 iterations is also possible. In general, the nonlinear inversion block should be chosen such that $H_{11}, H_{12}, H_{21}, H_{22}$ can be identified in the most effective manner. One popular approach for to minimize the cost function $C$ is to use the least mean square (LMS) algorithm, by which the coefficients of $H_{11}, H_{12}, H_{21}, H_{22}$ are updated as:

$$H_{11}^{n+1}(j) = H_{11}^n(j) + \mu_1 e(n) \frac{\partial d(n)}{\partial H_{11}(j)}, \quad (46)$$

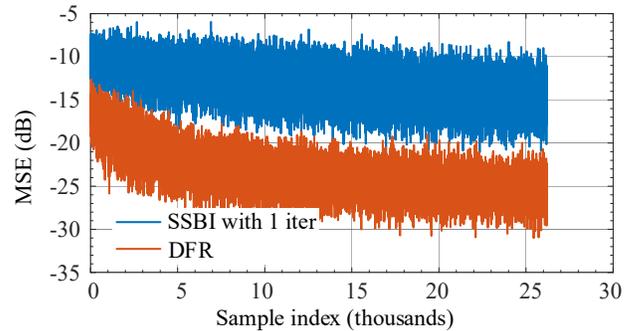

Fig. 16. Mean square error at the output of the proposed digital circuit in Fig. 15 with the nonlinear inversion block chosen as shown in Fig. 16. The training sequence is 100 Gbaud 16 QAM, both Tx and Rx has a 3-dB bandwidth of 35 GHz and their responses are modelled as 2-nd order Gaussian filter. The LOSPR is 13 dB

$$H_{12}^{n+1}(j) = H_{12}^n(j) + \mu_2 e(n) \frac{\partial d(n)}{\partial H_{12}(j)}, \quad (47)$$

$$H_{21}^{n+1}(j) = H_{21}^n(j) + \mu_1 e(n) \frac{\partial d(n)}{\partial H_{21}(j)}, \quad (48)$$

$$H_{22}^{n+1}(j) = H_{22}^n(j) + \mu_2 e(n) \frac{\partial d(n)}{\partial H_{22}(j)}, \quad (49)$$

where $j = 1,2,..L$ and $L$ is the filter length (we assume $H_{11}, H_{12}, H_{21}, H_{22}$ have the same length); $\mu_1$ and $\mu_2$ are the conversion parameters. The exact update rules for $H_{11}, H_{12}, H_{21}, H_{22}$ for considered nonlinear inversion blocks in Fig. 16 are shown in Appendices II and III. In general, the update rules when Eq. 18 and 19 (DFR technique) are used as the nonlinear inversion block is much more complicated than when the iterative SSBI cancellation scheme with 1 iteration is chosen as the nonlinear inversion block. On the other hand, as shown in Fig. 16, a much better performance can be achieved using the DFR technique where the MSE (defined by Eq. 44 normalized to the signal power) can be suppressed to ~ -25 dB. This MSE indicates that both the Tx and Rx responses have been identified with a high accuracy. Fig. 17 confirms this argument by showing the true and estimated Tx and Rx responses for the considered simulated system (100 Gbaud 16 QAM with a LOSPR of 13 dB, both Tx and Rx has a 3-dB bandwidth of 35 GHz and their responses are modelled as 2-nd order Gaussian filters).

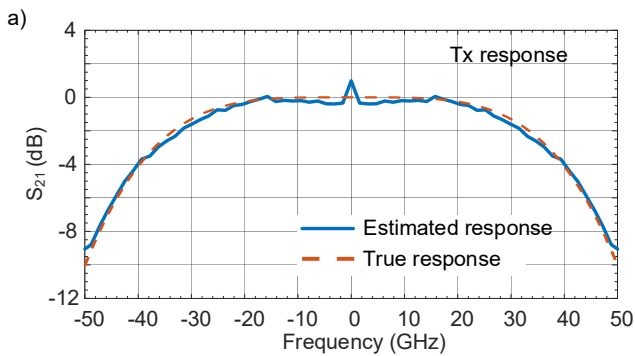

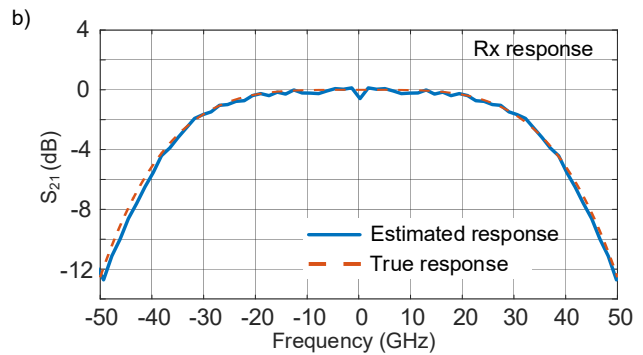

Fig. 17. The true and estimated Tx and Rx responses obtained using the proposed digital circuit (Fig. 14)

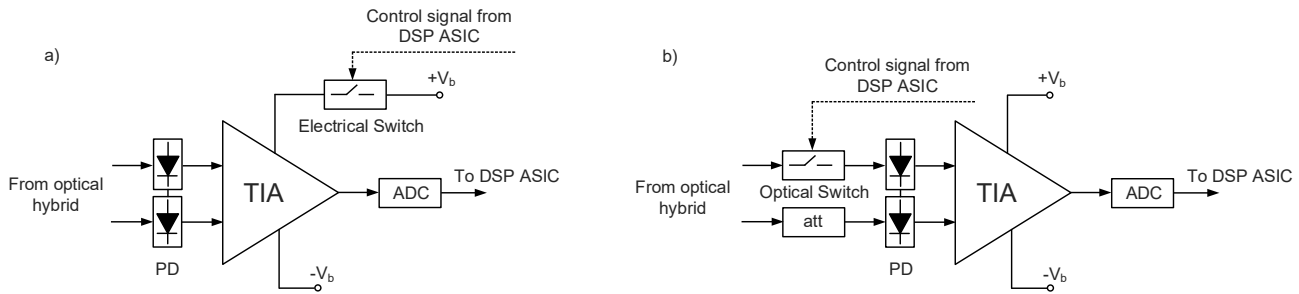

Fig. 18. Two possible approaches for turning a BR into a SER for the calibration purposes. a) – using an electrical switch for terminating an electrical tributary within the TIA; b) – using an optical switch for terminating an arm of the BPD.

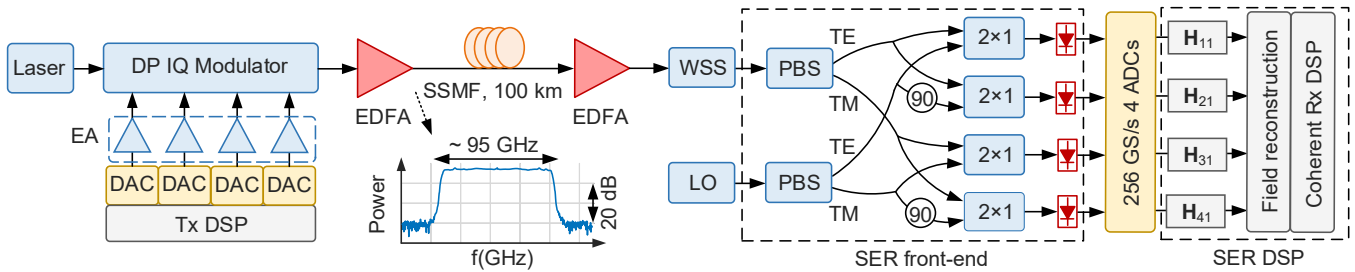

Fig. 19. Experimental setup for 90 Gbaud PCS-64QAM transmission over 100 km using a self-calibrated SER; PBS – polarization beam splitter, WSS – wavelength selective switch

In simulation we observe that the performance of the proposed technique depends on critical system's parameters such as the LOSPR value and ASE noise. There should be a certain limitation on the range where the algorithm can converge due to its nonlinear structure. This parametric study, however, is out of the scope of this paper and will be addressed in future studies. With some hardware modification, the idea of identifying both the Tx and Rx responses using a single training sequence can be extended for conventional BR. One viable approach is to make a modification which can turn a conventional BR into a SER for the calibration purposes. This can be achieved, for example, with two approaches shown in Fig. 18. The feasibility and practicality of these two approaches are a subject of future investigations.

## V. TRANSMISSION EXPERIMENT

In this section, we present a proof of concept experiment with 90 Gbaud PCS-64 QAM transmission over 100 km of SSMF which was first reported in [26]. The experimental setup for 90 Gbaud PCS-64 QAM with a SER is shown in Fig. 19. At the Tx, 90 Gbaud DP PCS-64 QAM signal with the entropy of 5.6 bits/symbol/polarization was generated using an RRC filter with a roll-off factor of 5%. After that, pre-emphasis was performed to approximately compensate for the overall Tx response, including the responses of DAC, RF driver and DP-I/Q modulator. The generated signal was then loaded into the memories of 4 CMOS DACs running at 120 GS/s. After optical modulation, the optical signal was amplified and launched into a single span of 100 km of SSMF. At the Rx, the signal was amplified and then passed through a WSS with an opened window of ∼ 125 GHz for ASE noise rejection. Next, the signal was detected by a SER front-end with 4 single-ended PD having 3-dB bandwidth of ∼ 70 GHz. In this experiment, we form a

SER from a conventional BR by terminating 4 optical paths to 4 balanced PDs. Finally, the signal was digitized by a 4-channel 256 GS/s 110 GHz real-time oscilloscope for offline signal processing.

Offline signal processing first includes resampling to 2 samples/symbol, O/E front-end characterization using a 16 QAM training sequence and the proposed adaptive algorithm for obtaining 4 FIR filters $H_{11}, H_{21}, H_{31}, H_{41}$. Next, field reconstruction was performed at 2 samples per symbol, using filed reconstruction technique discussed in previous sections. After that, the signal was fed into a conventional coherent DSP for symbol detection, BER counting, GMI calculation and 2D net information rate (IR) calculation using SD-LDPC decoding with code rate optimization [27].

We applied the self-calibration scheme shown in Fig. 14 where the nonlinear inversion block is based on the DFR technique for measuring the Rx responses. In our setup, due to the flat frequency response of the real-time scope (up to 100 GHz), the Rx non-ideal response is mainly due to the PDs. The training sequence was a 16QAM signal at the same baudrate in a single polarization. The algorithm was performed at 2 samples per symbol and the lengths of FIR filters were all set to 33. Using a short filter length reduces the implementation complexity. In addition, if the filter length is too long it could lead to performance instabilities due to the nonlinear structure of the proposed scheme. This issue will be addressed in future studies. The MSEs (the normalized cost function $C$) are shown in Fig. 20a for x and y polarizations. One can see that convergence is achieved after ∼ $10^5$ samples (at 180 GSa/s). At convergence, the MSEs for both polarizations are ∼ -18 dB, which indicates that the Rx response has been measured with good accuracy. The converged $H_{11}, H_{21}, H_{31}, H_{41}$ filters are shown in Fig. 20b. These filters were then used for front-end correction before

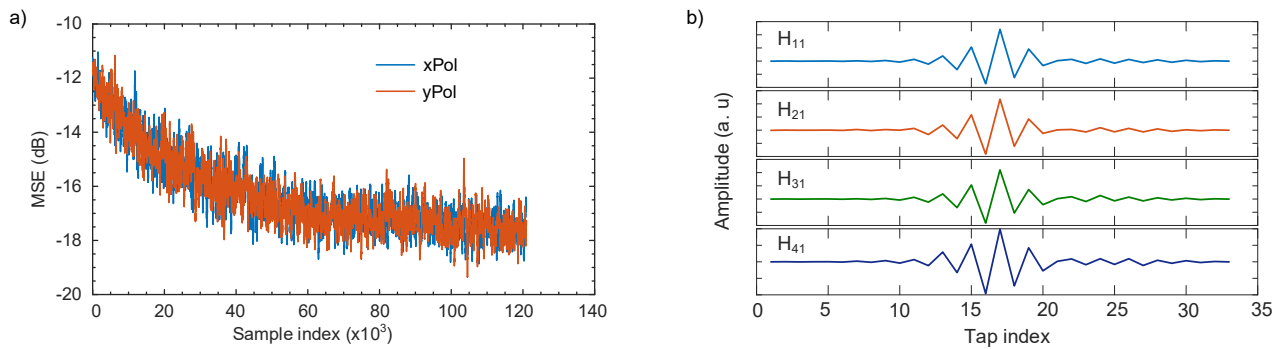

Fig. 20a) – The evolutions of MSE of the proposed self-calibration scheme (Fig. 14) for x and y polarizations; b) – Converged $H_{11}, H_{21}, H_{31}, H_{41}$ filters which approximate the impulse responses of the Rx (4 PDs)

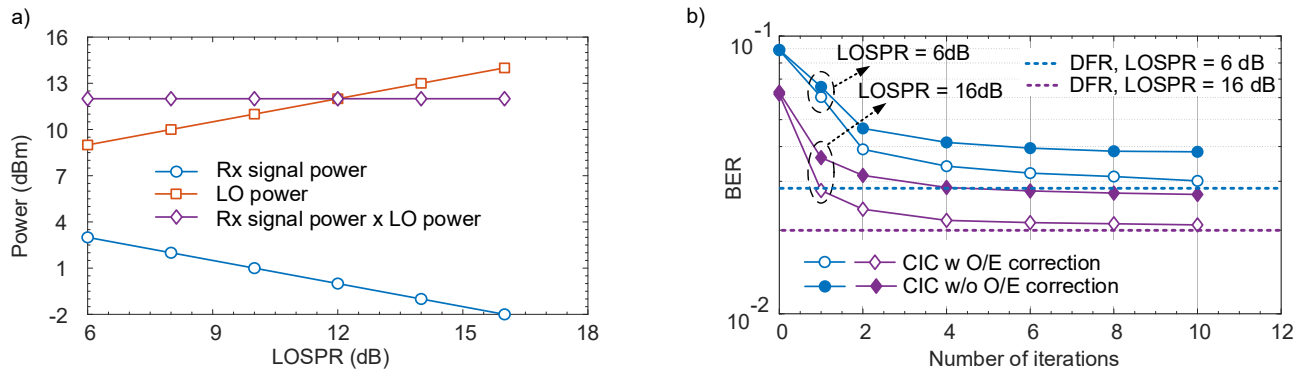

Fig. 21a) – Adjusting the LOSPR by varying both the Rx signal power and the LO power; b) – Performance of CIC technique versus number of iterations for LOSPR of 6 dB and 16 dB

optical filed reconstruction, channel equalization and symbol detection.

LOSPR is an important parameter of the investigated system. To vary the LOSPR, we vary the received signal power and the LO power simultaneously such that their product remains unchanged as shown in Fig. 21a. In this case, the detected signal swing on the real-time scope was maintained and a fair comparison for different values of LOSPR could be made.

The performances of the CIC technique with and without O/E front-end correction (using $H_{11}, H_{21}, H_{31}, H_{41}$) in B2B case at full OSNR are shown in Fig. 21b for LOSPR of 6 dB and 16 dB. Herein, the optimum clipping levels are 5 dB and 11 dB for LOSPR of 6 dB and 16 dB respectively. One can note that increasing the number of iterations improves the system performance continuously which confirms the effectiveness of the proposed clipping technique. However, after 6 iterations the performance improvement is negligible. Significant performance discrepancy can be observed between the case of with and without O/E front-end correction even through the Rx has a sufficiently wide bandwidth (the attenuation at ~ 50 GHz is only ~ 1.5 dB). This clearly indicates that O/E front-end correction is necessary for SER for achieving the best performance. We also show the performance of DFR technique in the case of with O/E front-end correction in Fig. 21b. One can note that DFR performs slightly better than the CIC technique due to the wide Rx bandwidth as discussed in the section III. We also tested the GD technique for the investigated system and found that it performs similar to the CIC technique. However, similar to what was observed in simulation, GD technique converges quite slowly, and it does not show advantage over the DFR technique when the Rx has sufficient bandwidth. Due to the limited added value for the investigated system, we will not discuss further the GD technique in this paper.

The system performance in B2B is summarized in Fig. 22a. Herein, we consider various configurations including the conventional SER without SSBI mitigation, SER with DFR and the conventional BR. One can note that the conventional SER performs poorly even when the LOSPR was set to 16 dB. On the other hand, for SER with DFR technique, a BER bellow the common soft-decision FEC threshold of 0.04 could already be achieved. When the LOSPR was decreased to 10 dB, SER with DFR shows a comparable performance with the BR. This indicates that the SSBI has been effectively mitigated by the DFR technique. Fig. 22b shows the OSNR penalty (at the BER of 0.04) of the SER with DFR technique in comparison with the conventional BR with 16 dB of LOSPR. One can note that when O/E front-end correction is not applied, the OSNR penalty can be as high as 2 dB when the LOSPR was set to 10 dB. On the other hand, when O/E front-end correction is applied, the OSNR penalty can be reduced to ~ 0.5 dB. This clearly indicates the effectiveness and necessity of the proposed self-calibration scheme for SER. A SER can only produce a competitive performance when an effective O/E front-end correction scheme is available for it.

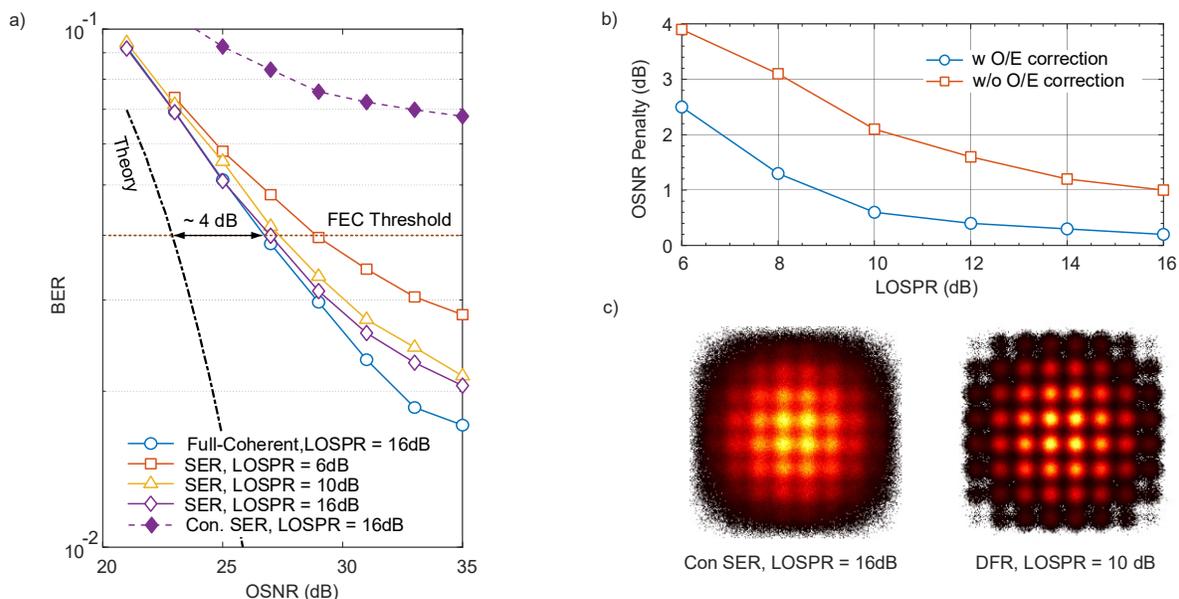

Fig. 22a) – BER versus OSNR for SER with DFR method, conventional SER without SSBI mitigation and the BR (full coherent ); b) – OSNR penalty to the full coherent Rx for the DFR technique; c) – Constellations for conv. SER and self-calibrated SER with DFR method

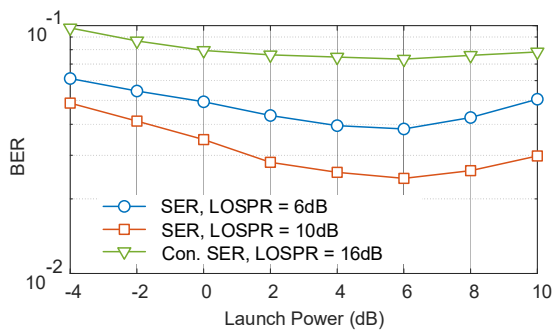

Fig. 23. BER vs the launch power for the transmission over 100 km

For the transmission over 100 km of SSMF, we first optimize the launched power as shown in Fig. 23. For the SER with DFR technique, the optimum launched power was 6 dBm, regardless of the LO power. At the optimum launched power, the 2D information rate (IR) achieved using the CIC technique with various number of iterations is shown in Fig. 24a. The case of 0 iteration indicates the performance of the conventional SER.

We consider here both the GMI and net bitrate obtained after FEC decoding using a family of SC-LDPC codes [27]. One can note that the CIC scheme increases the 2D net IR by ∼ 0.9 bit/symbol compared to the conventional SER (indicated by the case of 0 iteration). Fig. 24b indicates that for SER, the information rate increases with increasing LOSPR, as expected. However, due to the effectiveness of SSBI mitigation, a LOSPR of ∼ 12 dB would be enough for achieving the best performance, even for high spectral efficiency (SE) systems. This clearly indicates the effectiveness and attractiveness of algorithms presented in this paper. Overall, if we consider the DFR and a low complexity CIC technique with only 4 iterations, a 2D net IR of 4.9 bit/symbol can easily be achieved, leading to a net data rate of 882 Gb/s. This result indicates that SER can be an attractive Rx option for 800 ZR applications.

## VI. CONCLUSION

In this paper, we have shown that the SSBI in single-ended coherent receivers can be effectively mitigated in the digital domain using serval techniques, namely DFR, CIC and DG. In addition, the Rx response of a SER can also be measured using a training sequence from the Tx. The combination of SSBI

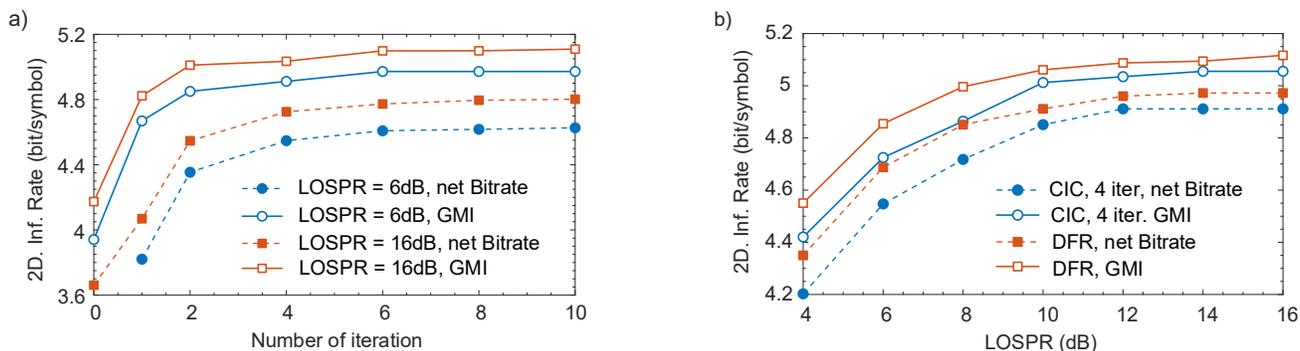

Fig. 24a) – 2D IR versus number of iterations for CIC scheme (0 iteration is the conventional SER scheme); b) – 2D IR versus LOSPR for SER with DFR and CIC with 4 iterations.

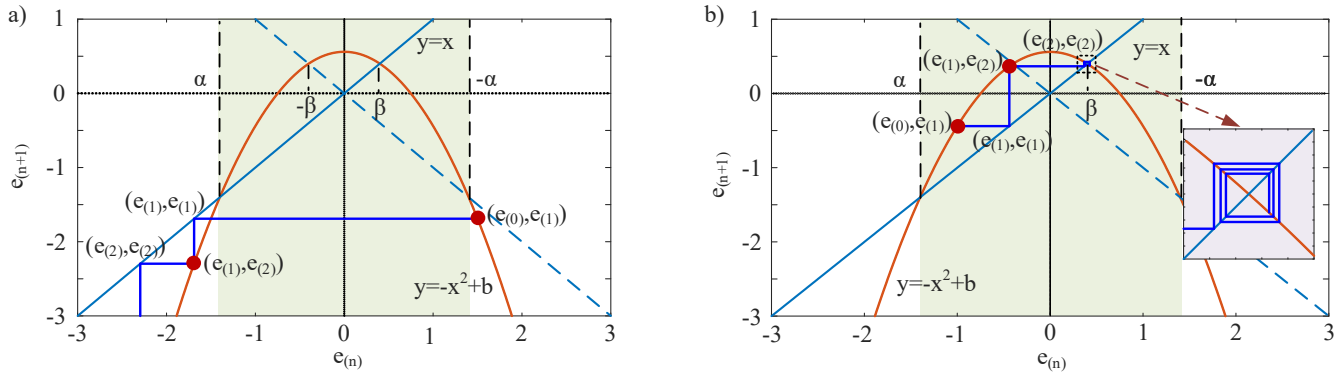

Fig. 25a – A Cartesian coordinate plane showing an exemplary evolution of $(e_{(n)}, e_{(n+1)})$ when $|e_{(0)}| > |\alpha|$, indicating that $e_{(n)} \to -\infty$ when $n \to \infty$; b) – A Cartesian coordinate plane showing an exemplary evolution of $(e_{(n)}, e_{(n+1)})$ when $|e_{(0)}| < |\alpha|$ and $b < 0.75$, indicating that $e_{(n)} \to \beta$ when $n \to \infty$

cancellation and Rx self-calibration can close the performance gap between a SER and a conventional BR. As a result, due to the simpler Rx architecture and lower cost, SER can be very attractive for low-cost pluggable coherent transceiver market. Using the developed techniques, we have demonstrated a 882 Gb/s transmission over 100 km of SSMF using a SER with a low LOSPR of 12 dB. The obtained results indicate that SER can provide a comparable performance compared to the conventional BR and thus is an attractive option for 800 ZR applications.

## APPENDIX I

To study the convergence behavior of (31), we make the change of variable as:

$$e_{(n)} = 2\Delta\bar{I}_{(n)} + \bar{I} + \overline{Q}, \tag{50}$$

Then, one can then verify from Eq. 31 that

$$e_{(n+1)} = -e_{(n)}^2 + b, \tag{51}$$

where $b = (\bar{I} + \overline{Q})^2 + \bar{I} + \overline{Q}$.

The recurrence relation (51) is a form of the "quadratic map" [15]. The quadratic map is capable of very complicated behavior. While some quadratic maps are solvable in closed form (for example, there are three solvable cases in the "logistic map" [16]), most are not. The recurrence relation (51) is also not solvable in the general form (for arbitrary values of $b$ and $e_{(0)}$. In the following, we will discuss some important features and behaviors of the recurrence relation (51).

An important boundary condition is given by:

$$b = \left(\bar{I} + \overline{Q} + \frac{1}{2}\right)^2 - \frac{1}{4} \ge -\frac{1}{4}.$$

If the sequence $e_{(n)}$ converges to $x$ when $n \to \infty$, we must have:

$$x = -x^2 + b, \tag{52}$$

which always has two real-valued roots:

$$\alpha = -\frac{1}{2} - \frac{1}{2}\sqrt{1 + 4b} \text{ and } \beta = -\frac{1}{2} + \frac{1}{2}\sqrt{1 + 4b}. \tag{53}$$

In the vicinity of $\alpha$, e. g., $|e_{(n)} - \alpha| < \varepsilon$, with $\varepsilon$ being a small positive number, one can verify that:

$$|e_{(n+1)} - \alpha| = |e_{(n)} - \alpha||e_{(n)} + \alpha| \ge |e_{(n)} - \alpha|, \tag{54}$$

which means that $e_{(n)}$ never converges to $\alpha$ except if $e_{(0)} = \alpha$. Similarly, in the vicinity of $\beta$, the convergence of $e_{(n)}$ implies

$$|2\beta| = \left|-1 + \sqrt{1 + 4b}\right| \le 1 \text{ or } b \le \frac{3}{4}, \tag{55}$$

or equivalently,

$$-1.5 \le \bar{I} + \overline{Q} \le 0.5. \tag{56}$$

The condition (55) means that if $b > 3/4$, then $e_{(n)}$ does not converge to a single value, so the condition (33) is not met, resulting in a nonzero reconstruction error.

Now, assume that $e_{(n)}$ converges to $\beta$ when $b \le 0.75$, then from (34) one cans show that:

$$\Delta\bar{I}_{(n)} \to -\frac{1}{4} + \frac{1}{4}\left|1 + 2\bar{I} + 2\overline{Q}\right| - \frac{\bar{I} + \overline{Q}}{2} \text{ when } n \to \infty, \tag{57}$$

which is equivalent to

$$\Delta\bar{I}_{(n)} \to 0 \text{ if } |\bar{I} + \overline{Q}| \le 0.5, \tag{58}$$

and

$$\Delta\bar{I}_{(n)} \to -(\bar{I} + \overline{Q} + 0.5) \text{ if } -1.5 \le \bar{I} + \overline{Q} < -0.5. \tag{59}$$

Then we arrive at the first important conclusion that $\Delta\bar{I}_{(n)} \to 0$ if $|\bar{I} + \overline{Q}| < 0.5$ and $e_{(n)}$ converges to $\beta$.

On the other hand, when $|e_{(n)}| > |\alpha|$, one can easily show that $e_{(n+1)} < \alpha$ and for all $m > n + 1$ we have:

$$\alpha - e_{(m+1)} = -(\alpha - e_{(m)})(\alpha + e_{(m)}) > (\alpha - e_{(m)}) > 0, \tag{60}$$

which implies that $e_{(n)} \to -\infty$ when $n \to \infty$. This case is illustrated in the Cartesian coordinate plane of $(e_{(n)}, e_{(n+1)})$ in Fig. 25a, where $e_{(n)}$ monotonically decreases to $-\infty$. This shows that if there exists any index $n$ such that $|e_{(n)}| > |\alpha|$, $e_{(n)}$ will diverge to $-\infty$.

Now we should find the condition that $e_{(n)}$ is confined within the interval $[\alpha, -\alpha]$. This condition implies that:

$$e_{(0)} = 2\left(\bar{I}^2 + \overline{Q}^2 - \frac{P}{4A^2}\right) + \bar{I} + \overline{Q} \le |\alpha|, \tag{61}$$

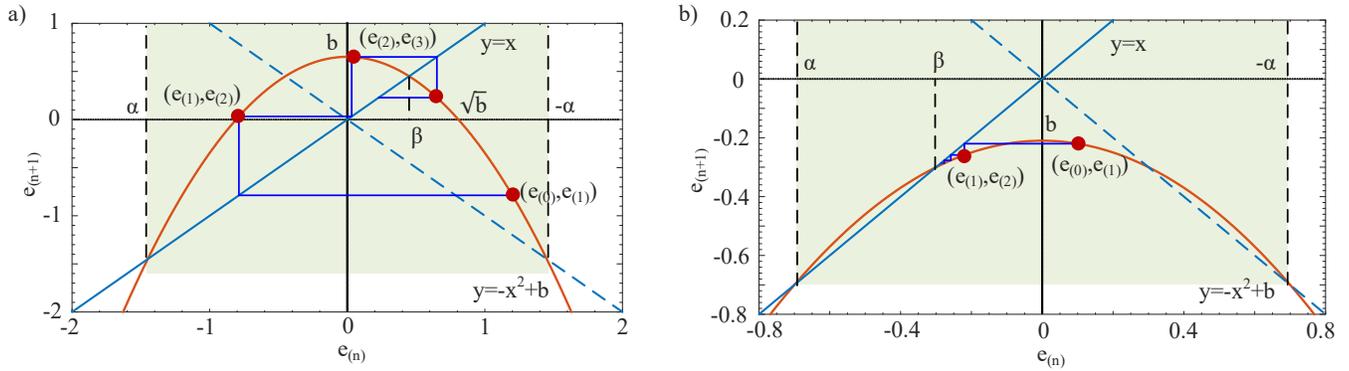

Fig. 26a) – A Cartesian coordinate plane showing an evolution of $(e_{(n)}, e_{(n+1)})$ when $0 \leq b \leq 0.75$ b) – A Cartesian coordinate plane showing an evolution of $(e_{(n)}, e_{(n+1)})$ when $-0.25 \leq < 0$

and

$$|-x^2 + b| \leq |\alpha| \text{ if } |x| \leq |\alpha|. \tag{62}$$

One can note that the condition (61) is the same as (34).
As $|-\alpha^2 + b| = |\alpha|$, the condition (62) simply implies that $b \leq |\alpha|$ or

$$b \leq \frac{1}{2} + \frac{1}{2}\sqrt{1 + 4b} \text{ or } b \leq 2. \tag{63}$$

The condition (63) is equivalent to:

$$-2 \leq \overline{I} + \overline{Q} \leq 1. \tag{64}$$

Now we shall show that if the condition (61) is satisfied and $e_{(0)} \neq |\alpha|$ then $e_{(n)}$ converges to $\beta$ when $b \leq 0.75$.
We consider two cases:

*a)* if $0 \leq b \leq 0.75$:

One can easily show that

$$\beta = -\frac{1}{2} + \frac{1}{2}\sqrt{1 + 4b} \geq 0. \tag{65}$$

First, one can easily prove that there exists an index $n$ such that $0 \leq e_{(n)} \leq \sqrt{b}$ through the following observations: *i)* If $\sqrt{b} < e_{(n)} < |\alpha|$ then $e_{(n+1)} = -e_{(n)}^2 + b < 0$. *ii)* If $-\sqrt{b} \leq e_{(n)} <$

$0$, then $0 < e_{(n+1)} < b < \sqrt{b}$; *iii)* If $\alpha \leq e_{(n)} \leq -\sqrt{b}$ then $e_{(n+1)} = -e_{(n)}^2 + b > e_{(n)}$ thus $e_{(n)}$ keeps increasing until it falls into the interval $[-\sqrt{b}, 0]$ and then $[0, \sqrt{b}]$. The illustration in Fig. 26a shows how $e_{(n)}$ should enter the interval $[0, \sqrt{b}]$ at some point.
Second, if $0 \leq e_{(n)} \leq \sqrt{b}$, then $e_{(n+1)} = -e_{(n)}^2 + b > 0$ and $e_{(n+1)} < b < \sqrt{b}$. This means that $0 \leq e_{(m)} \leq \sqrt{b}$ for all $m > n$. Now considering an index $m > n$ we have:

$$|e_{(m+1)} - \beta| = |\beta - e_{(m)}|(\beta + e_{(m)}) \tag{66}$$

$$= |e_{(m-1)} - \beta|(\beta + e_{(m)})(\beta + e_{(m-1)}). \tag{67}$$

We now will prove that $|e_{(m+1)} - \beta| \leq |e_{(m-1)} - \beta|$.
One can note that

$$(\beta + e_{(m)})(\beta + e_{(m-1)}) \leq \frac{1}{4}(2\beta + e_{(m)} + e_{(m-1)})^2$$

$$= \frac{1}{4}(2\beta - e_{(m-1)}^2 + e_{(m-1)} + b)^2$$

$$= \frac{1}{4}\left(2\beta + \frac{1}{4} - \left(e_{(m-1)} - \frac{1}{2}\right)^2 + b\right)^2$$

$$\leq \frac{1}{4}(2\beta + 0.25 + b)^2 \leq 1$$

because $0 \leq b \leq 0.75$.
The sign $=$ happens only when $e_{(m)} = e_{(m-1)} = 0.5$ and $b = 0.75$ which implies that $e_{(m)} = \beta$ already. If the sign $=$ does not happen, $|e_{(m+1)} - \beta| < |e_{(m-1)} - \beta|$ which guarantees that $|e_{(m+1)} - \beta|$ converges and in this case it must converge to 0. In other words, $e_{(n)}$ converges to $\beta$.

*a)* if $-0.25 \leq b < 0$:

This case is simpler and is illustrated in Fig. 26 where $\alpha < \beta < 0$. Using the same argument, we can show that an index $n$ exists such that $\alpha < e_{(n)} \leq 0$. Then we only need to consider two intervals. *i)* if $\alpha < e_{(n)} \leq \beta$, we have:

$$(e_{(n+1)} - \beta) = -(e_{(n)} - \beta)(\beta + e_{(n)}). \tag{68}$$

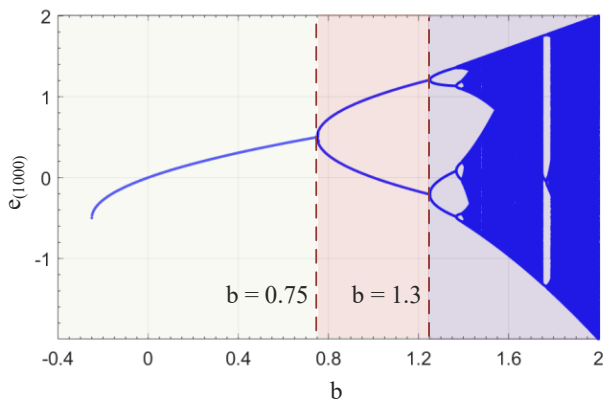

Fig. 27. Bifurcation diagram of $e_{(1000)}$ versus $b$ obtained from $10^5$ random samples of $e_{(0)}$ for each $b$, which satisfies the condition (61).

As $-1 = \alpha + \beta \leq (\beta + e_{(n)}) < 0$, one can verify that:

$$0 > (e_{(n+1)} - \beta) > (e_{(n)} - \beta). \qquad (69)$$

This guarantees that $e_{(n)} - \beta$ converges and thus $e_{(n)}$ must converge to $\beta$. *ii)* if $\beta < e_{(n)} \leq 0$, using the recurrence relation (66), we have $0 < e_{(n+1)} - \beta < 2|\beta|(e_{(n)} - \beta)$. As $|\beta| \leq 0.5$, $0 < e_{(n+1)} - \beta < (e_{(n)} - \beta)$. This also guarantees that $e_{(n)} - \beta$ converges and thus $e_{(n)}$ must converge to $\beta$.

When (61) is satisfied but $b > 2$ (i.e., $\overline{I} + \overline{Q} > -2$ or $\overline{I} + \overline{Q} > 1$), there is no guarantee that $e_{(n)}$ is bounded. As $e_{(n)}$ does not converge to a single point in this case, it can either go to $-\infty$ or bounded within $[\alpha, -\alpha]$.

The last case to consider is when (61) is satisfied (and $e_{(0)} \neq |\alpha|$) and $0.75 < b \leq 2$. In this case, the recurrent sequence $e_{(n)}$ exhibits a complicated behavior. To study its behavior, we show a bifurcation diagram [17] for $e_{(1000)}$ as function of $b$ in Fig. 27, obtained with $10^5$ random values of $e_{(0)}$ which satisfies the condition (61) for each $b$. One can see that when $0.75 < b \leq 1.3$, $e_{(1000)}$ can have two distinct values; indeed, this phenomenon occurs since it *oscillates* between 2 different values depending on the initial value $e_{(0)}$. One should note that here $e_{(1000)}$ represents the behavior of $e_{(n)}$, $n \to \infty$. When $b > 1.3$, $e_{(1000)}$ can bounce between 4, 8 different values or even goes in a chaotic regime.

## Appendix II

Herein we derive the LMS update rules for $H_{11}, H_{12}, H_{21}, H_{22}$ based on Eq. 46 – Eq. 49 when the nonlinear inversion block is the iterative SSBI cancellation scheme with 1 iteration.

As shown in Fig. 15 and 16, we have:

$$I_1(n) = \overline{R}_1(n) \otimes H_{11}; \ Q_1(n) = \overline{R}_2 \otimes H_{21}, \qquad (71)$$

$$I_2(n) = I_1(n) - (I_1(n)^2 + Q_1(n)^2) \qquad (72)$$

$$Q_2(n) = Q_1(n) - (I_1(n)^2 + Q_1(n)^2) \qquad (73)$$

$$I_3(n) = I_2(n) \otimes H_{12}; \ Q_3(n) = Q_2(n) \otimes H_{22} \qquad (74)$$

$$d(n) = I_2^2(n) + I_2^2(n); \ e(n) = |s(n)|^2 - d(n) \qquad (75)$$

Then using Eq. 46 – Eq. 49, we can define the update rules for $H_{11}, H_{12}, H_{21}, H_{22}$ as:

$$H_{11}^{(n+1)} = H_{11}^{(n)} + \mu_1 e(n) \sum_{k=1}^{L} (I_3(n)H_{12}(k)(1 - I_1(n-k)) - Q_3(n)H_{22}(k)I_1(n-k))R_1(n-k), \quad (76)$$

$$H_{12}^{(n+1)} = H_{12}^{(n)} + \mu_2 e(n) I_3(n) I_2(n), \qquad (77)$$

$$H_{21}^{(n+1)} = H_{21}^{(n)} + \mu_1 e(n) \sum_{k=1}^{L} (Q_3(n)H_{22}(k)(1 - Q_1(n-k)) - I_3(n)H_{12}(k)Q_1(n-k))R_2(n-k) \quad (78)$$

$$H_{22}^{(n+1)} = H_{22}^{(n)} + \mu_2 e(n) Q_3(n) Q_2(n), \qquad (79)$$

where $j = 1, 2, \ldots L$, bold symbols denote vectors of length $L$, for example: $I_1(n) = [I_1(n), I_1(n-1), \ldots I_1(n - L + 1)]$; $H_{12}^{(n)} = [H_{12}^{(n)}(1), H_{12}^{(n)}(2), \ldots H_{12}^{(n)}(L)]$

## Appendix III

Herein we derive the LMS update rules for $H_{11}, H_{12}, H_{21}, H_{22}$ based on Eq.46 – Eq. 49 when the nonlinear inversion block is the Eq. 18 and 19. Herein, the relation between $(I_2(n), Q_2(n))$ and $(I_1(n), Q_1(n))$ is given by Eq. 18 and 19. Based on Eq. 56-49, we have:

$$H_{12}^{(n+1)} = H_{12}^{(n)} + \mu_2 e(n) I_3(n) I_2(n), \qquad (80)$$

$$H_{11}^{(n+1)}(j) = H_{11}^{(n)}(j) + \mu_1 e(n)\left(I_3(n)\frac{\partial I_3(n)}{\partial H_{11}(j)} + Q_3(n)\frac{\partial Q_3(n)}{\partial H_{11}(j)}\right)$$

$$\frac{\partial I_3(n)}{\partial H_{11}(j)} = (R_1(n-j)(\frac{1}{4A} + \frac{1}{4\sqrt{\Delta(n)}}(1 - \frac{1}{2A^2}(I_1(n) - Q_1(n)))) * H_{12}^{(n)} \quad (81)$$

$$\frac{\partial Q_3(n)}{\partial H_{11}(j)} = (R_1(n-j)(-\frac{1}{4A} + \frac{1}{4\sqrt{\Delta(n)}}(1 - \frac{1}{2A^2}(I_1(n) - Q_1(n)))) * H_{22}^{(n)}, \quad (82)$$

where $j = 1, 2, \ldots L$, bold symbols denote vectors of length $L$, and $*$ denotes the vector product.

The updating rules for $H_{21}, H_{22}$ can be derived in a similar manner.